\documentclass[11pt,a4paper]{article}
\usepackage[utf8]{inputenc}

\usepackage[T1]{fontenc} 
\usepackage{microtype} 
\usepackage[belowskip=-0pt,aboveskip=0pt]{caption}  
\usepackage[defblank]{paralist} 
\usepackage[noend]{algpseudocode}
\usepackage[numbers,square,sort&compress]{natbib}
\usepackage[obeyspaces]{url} 
\usepackage{algorithm}
\usepackage{amsfonts,amsmath,amssymb} 
\usepackage{datetime}
\usepackage{enumitem} \setlist[enumerate,1]{label=\arabic*.,} \newlist{inlinelist}{enumerate*}{1} \setlist*[inlinelist,1]{label=(\roman*),} \setlist[enumerate]{leftmargin=\parindent}
\usepackage{etoolbox}
\usepackage{graphicx}
\usepackage{multirow}
\usepackage{parskip} \parskip=1ex plus 5pt minus 1pt \parindent=4ex 
\usepackage{subcaption}
\usepackage{xcolor}
\usepackage{xspace}

\usepackage{hyperref}  

\algdef{SE}[SUBALG]{Indent}{EndIndent}{}{\algorithmicend\ }%
\algtext*{Indent}
\algtext*{EndIndent}    

\newcommand{\blfootnote}[1]{}

\newcommand{\HB}{\ensuremath{\rightarrow}}
\newtheorem{definition}{Definition}
\newtheorem{proposition}{Proposition}
\newtheorem{lemma}{Lemma}

\newcommand{\SYSTEM}{Cure} 

\begin{document}

\title{
  Distributed transactional reads: \\
  the strong, the quick, the fresh \& the impossible%
\thanks{Preprint from \href{http://2018.middleware-conference.org/}{{19th International Middleware Conference},
  {10--14 December 2018}, {Rennes, France}}}
}

\author{Alejandro Z. Tomsic \\
  {\small Sorbonne Université, Inria, LIP6, Paris}
  \and Manuel Bravo \\
  {\small IMDEA Software Institute, Madrid}%
  \thanks{Work partially done as a student at Université Catholique de Louvain and Universidade de Lisboa}
  \and Marc Shapiro \\
  {\small Sorbonne Université, Inria, LIP6, Paris}
}

\date{18 September 2018}

\maketitle

\begin{abstract}
  This paper studies the costs and trade-offs of providing transactional
  consistent reads in a distributed storage system.
  We identify the following dimensions:
  read consistency, read delay (latency), and data freshness.
  We show that there is a three-way trade-off between them, which can be
  summarised as follows:
  \begin{inparaenum}[\it (i)]
  \item
    it is not possible to ensure at the same time order-preserving
	(e.g., causally-consistent) or
    atomic reads, Minimal Delay, and maximal freshness; thus,
    reading data that is the most fresh without delay is possible only
    in a weakly-isolated mode;
  \item
    to ensure atomic or order-preserving reads at Minimal Delay
    imposes to read data from the past (not fresh);
  \item
    however, order-preserving minimal-delay reads can be fresher than atomic;
  \item
    reading atomic or order-preserving
    data at maximal freshness may block reads or writes indefinitely.
  \end{inparaenum}
  Our impossibility results hold independently of other features of the
  database, such as update semantics (totally ordered or not) or data
  model (structured or unstructured).
  Guided by these results, we modify an existing protocol to 
  ensure minimal-delay reads (at the cost of
  freshness) under atomic-visibility and causally-consistent semantics.
  Our experimental evaluation supports the theoretical results.
\end{abstract}

\section{Introduction}
This\blfootnote{\textsuperscript{*}Work partially done as student at Universit\'e 
Catholique de Louvain and Universidade de Lisboa}
paper studies the costs of reading data in a distributed,
transactional storage system.
In particular, we try to understand whether it is possible to  provide
strong read guarantees while ensuring both fast performance and
fresh data.
Intuitively, stronger guarantees will come with higher costs.
A recent paper from Facebook (whose performance is strongly
read-dominated) states:%
\emph{``stronger properties have the potential to improve user
  experience and simplify application-level programming [\ldots{} but] are
  provided through added communication and heavier-weight state management
  mechanisms, increasing latency [\ldots{}] This may lead to a worse user
  experience, potentially resulting in a net detriment''}
\cite{facebook-challenges-consistency}.
Is this wariness justified, i.e., is it inherently impossible to combine
fast reads with strong guarantees, or can the situation be improved by
better engineering?
This paper provides a formal and operational study of the costs and
trade-offs.
Our main finding is that there is a three-way trade-off between read
guarantees, read delay (and hence latency), and freshness, and that some
desirable combinations are impossible.

\begin{figure}[tp]
  \centering
  \includegraphics[width=.9\columnwidth]{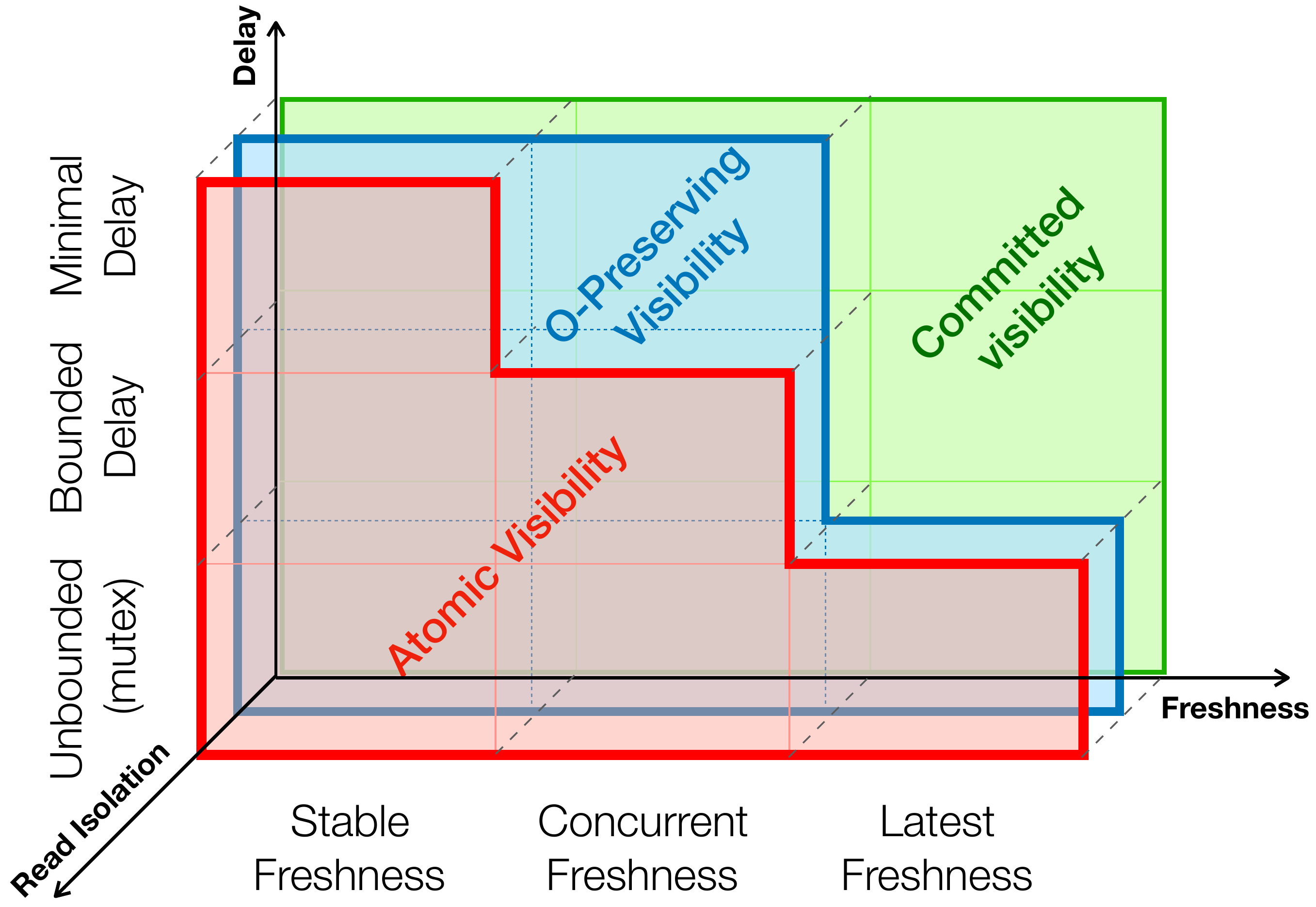}
  \caption{The three-way trade-off.  {\rm The boxed areas represent possible
    guarantee{\slash}read delay{\slash}freshness combinations.
    Upwards and right is better performance; guarantees are stronger
    from back to the front planes.}}
  \label{fig:3-way-tradeoff}
\end{figure}

It is well known that non-serialisable models, such as Snapshot
Isolation \cite{syn:bd:1759} or Highly-Available Transactions
\cite{db:syn:1751}, can improve availability and performance.
Therefore, in this paper, we do not necessarily assume that updates
are totally ordered.%
\footnote{
  Enforcing a total order of transactional updates enables Consistency under
  Partition (CP); but, conversely, Availability under Partition
  (AP) requires accepting concurrent updates \cite{syn:formel:sh190}.
}
Furthermore, we allow weakening the read guarantees: in addition to
Atomic Visibility, offered by
classical transactional models, we also consider (the weaker) Order-Preserving
Visibility, enforced by recent  causally-consistent
systems \cite{rep:syn:1662}, and (the weakest)
Committed Visibility, offered by Read Committed isolation \cite{ansi1992}.

Finally, we consider the \emph{freshness} dimension, because (as we
show) decreasing the read delay sometimes forces to read a version of
the data that is not the most recent.
Reading outdated data further incurs multi-version overhead
and, under strong consistency, increases aborts and reduces
throughput \cite{rep:syn:sh156}.

Figure~\ref{fig:3-way-tradeoff} illustrates the three-way trade-off
between the guarantees, delay, and freshness of transactional reads.
For instance, under Order-Preserving and Atomic Visibility, it is
possible to read with no extra delay (compared to a non-transactional
system), but then the freshest data is not accessible.
Transactions that access the freshest
data with no extra delay are only possible under Committed Visibility.
However, we show that
minimal-delay Order-Preserving reads allow observing updates of concurrently-committed
transactions.
As we will see in our evaluation section, this allows for
a significant freshness improvement over Atomic
Visibility, which forces reading data that was
stable (written and acknowledged) before the transaction sends its reads.
If, on the other hand, the application requires the freshest data, under
either Atomic or Order-Preserving Visibility, this is possible only under a
protocol where reads and writes are mutually exclusive, e.g., a read
might be delayed (blocked, or in a retry loop) indefinitely by
writes,  or vice-versa.


This work includes the following contributions:%

\begin{itemize}[leftmargin=*]
\item
  A formal study of the trade-offs
  between the read guarantees, delay, and
  freshness of transactional reads.
  We prove which desirable combinations are possible and which are not.
\item
  Guided by the results of our analysis, we modify an existing
  system that exhibits delays and derive three novel minimal-delay protocols.
  Each protocol trades freshness for isolation differently. AV provides Atomic Visibility and
  Stable Freshness, OP provides Order-Preserving Visibility and Concurrent Freshness,
  and CV Committed Visibility and Latest Freshness.
  We provide detailed protocol design, including pseudo-code.
\item
  An evaluation of these protocols to empirically validate our 
  theoretical results.
  In our measurements, we compare the introduced protocols
  to the protocol from which they derive. 
  Our minimal-delay protocols exhibit similar latency.
  CV always observes the most recent data,
  whereas freshness degrades negligibly for OP,
  and severely under AV.
\end{itemize}

\section{Model and Requirements}

In a typical distributed application deployment, we are interested in
the database (data storage) tier.
The database is distributed, i.e., data is partitioned
across many servers within a data centre.
Each partition is possibly replicated across several data centres.
Clients, representing application logic, access data by contacting the
appropriate servers within the data centre.
To increase performance, a multi-read{\slash}multi-write interface
enables the application to access multiple database servers at once in
parallel.
multi-write operations respectively.

\subsection{Transactions}
\label{sec:transactions}

The application consists of \emph{transactions}.
A transaction consists of any number of reads, followed by any number of
writes, and terminated by an abort (writes have no effect) or a commit
(writes modify the store; in what follows, we consider only committed
transactions).
The transaction groups together low-level storage operations into a
higher-level abstraction, with properties that help developers reason
about application behaviour.%
\footnote{
  A model that does not support transactions is identical to one where
  each individual read or write operation is wrapped in a transaction
  that commits immediately.
}
These are often summarised as the ACID properties (Atomicity,
Consistency, Isolation, and Durability).

\emph{Atomicity} ensures that, at any point in time, either all of a
transaction's writes are in the store, or none of them is.
This guarantee is essential for ensuring some common data invariants,
such as equality or complementarity between two data items (e.g.,
 the symmetry of the friendship or the like relationship in a social
 network application \cite{tao}).
 They are also instrumental to keep materialised views,
e.g., when adding a friend to a friends list, by updating
a friends count instead of
computing this count on reads \cite{ramp}.
%
\emph{Consistency} is the requirement that each of the
application's transactions individually transitions the database from a
valid state to another valid state.
\emph{Durability} means that later transactions will observe the effect
of this transaction after it commits.
\emph{Isolation} characterises non-interference between concurrent
transactions.

In the interest of performance and availability, we do not assume strong
consistency or isolation (e.g., serialisability). 
Specifically, neither writes nor reads are necessarily totally ordered,
and we consider several read guarantees.
Interestingly, our results are independent of the write
model, e.g.,  totally ordered or not. Atomicity and durability will
be taken for granted in the rest of this paper.

For simplicity we assume that a transaction reads a data item at most
once (and similarly for writes).
The set of item states read by a transaction is called its
\emph{snapshot}.
Our study distinguishes some important properties of a
snapshot, explained in the next few sections: snapshot guarantees,
delay, and freshness.

\subsection{Snapshot guarantees}
\label{sec-apguarantees}

\emph{Snapshot guarantees} constrain the states of the data items
that can be accessed by a given snapshot.
The stronger guarantees provide higher isolation, and thus facilitate
reasoning by the application developer.
As we shall see, the weaker ones enable better performance along the
freshness and delay metrics.
We distinguish three levels, which will be defined formally
later (in Section~\ref{sec-notation}): Committed, 
Order-Preserving, and Atomic Visibility.

  At the weakest level, \textbf{Committed Visibility},
  a snapshot may
  include any updates that have been committed.
  As it sets no constraints between items, it allows many 
  anomalies.
  Committed Visibility is equivalent to the read guarantees of Read
  Committed isolation \cite{ansi1992}.

  Recently, systems have proposed causally-consistent-snapshot reads
  \cite{rep:syn:1662}, which strengthen Committed Visibility by ensuring
  there are no happens-before-order \cite{syn:mat:1025} anomalies
  among the item states included in the snapshot.
  We consider \textbf{Order-Preserving Visibility}, 
  a generalisation of this property.
  Order-Preserving Visibility 
  ensures that the snapshot preserves \emph{some} (partial or total)
  order relation $O$.  
  $O$ might be the (partial) happens-before order, or the total order
  of updates in the context of a strong isolation criterion such as
  Serialisability or Snapshot Isolation \cite{bernstein, syn:bd:1759}.
  
   Order-Preserving Visibility ensures that transactions do not observe
gaps in a prescribed (partial or total) order relation.%
{%
  \newcommand{\photos}{\mathit{photos}}
  \newcommand{\acl}{\mathit{acl}}
  Consider, in a social network, the data items $\photos$ and $\acl$
  representing user Alice's photo album and the associated permissions.
  The set of their states (initially $\photos_{0}$ and $\acl_{0}$)
  follows a given order $\HB$.
  Alice changes the permissions of her photo album from public to
  private (new state $\acl_{1}$), then adds private photos to the album
  (state $\photos_{2}$).
  Thus, $\acl_{0} \HB \acl_{1} \HB \photos_{2}$. Unlike Committed
  Visibility, Order-Preserving Visibility disallows the situation where Bob would
  observe the old permissions ($\acl_{0}$) along with the new photos
  ($\photos_{2}$), missing out on the restricted permissions
  ($\acl_{1}$).
}
This pattern, where the application enforces
a relation between two data items by issuing updates in a particular
order, is typical of security invariants \cite{syn:formel:sh190}.
It also helps to preserve referential integrity (create
an object before referring to it, and destroy references before deleting
the referenced object).

\textbf{Atomic Visibility} is the strongest.
  It is order-preserving, and
  additionally disallows the ``read skew'' \cite{syn:bd:1759} 
  (also known as broken reads \cite{ramp}) phenomenon: if
  the transaction reads some data item written by another transaction, then
  it must observe all  updates written by that transaction (unless
  overwritten by a later transaction).
  In a system where atomic updates ensure
  Alice is in Bob's friends list if and only if Bob is in Alice's,
  Atomic Visibility ensures a transaction will observe both updates,
  or none.
  It also ensures observing materialised views consistently \cite{ramp}; 
  for instance, observing that the cardinality of a friends list 
  updated atomically with a friends count match.

  Atomic Visibility is the read guarantee provided by 
  Serialisability, Cursor Stability and Repeatable Reads ANSI
  isolation \cite{ansi1992}, of
  every strongly-consistent (totally-ordered) criteria
  (e.g., Parallel Snapshot Isolation \cite{rep:syn:1661},
  Non-Monotonic Snapshot Isolation \cite{rep:syn:sh156}, Snapshot
  Isolation \cite{syn:bd:1759}, Update Serialisability
  \cite{update-serializability-locking}, and Strict
  Serialisability \cite{external}), and the weakly-consistent
  Transactional Causal Consistency \cite{rep:pro:sh182, syn:rep:1708}, 
  and Read Atomicity \cite{ramp}.
  
\subsection{Delay}
\label{sec-low-delay}

Perceived low latency keeps users engaged and
directly affects revenue \cite{shopzilla, datauseful, latencyimpact}.
Furthermore, read latency is an important performance metric
for services that are heavily read-dominated, such as social
networks.
For instance, serving a Facebook page requires several rounds, where
each round reads many items, and what is read in one round depends on the
results of the previous ones; this amounts to tens of rounds and
thousands of items for a single page \cite{tao}.
For read-intensive applications, it is paramount to avoid read delays
\cite{facebook-challenges-consistency}.
Under existing systems, servers often delay a response to a read 
request. 
Read delay scenarios include wating for a lock to be released, for
physical clocks to catch up, or for a 
protocol (such as Two-Phase Commit \cite{twophasecommit}) to finish execution.

The fastest protocol exhibits \textbf{Minimal Delay}:
distributed reads can address multiple servers in parallel,
and any one server always responds with commited data 
\emph{immediately}, i.e., in a single round-trip, without
blocking or coordinating with other servers.
Intuitively, this design makes it difficult to ensure strong snapshot
guarantees.
Examples include Linkedin's Espresso \cite{linkedin-espresso} and
Facebook's Tao \cite{tao}.
We will consider Minimal-Delay as our baseline, and
will characterise protocols by estimating the \emph{added delay}
above this baseline.

\textbf{Bounded Delay} means that parallel reads are not supported, that a
bounded number of retry round-trips may occur to read from a server,
and{\slash}or that a server may block for a bounded amount of time before
replying to a read request.
An example is Eiger, which requires a maximum of three round-trips
to storage servers to read from a snapshot that preserves
Atomic Visibility under happens-before order
\cite{syn:rep:1708}.

\textbf{Mutex reads{\slash}writes} means that a read might be delayed
indefinitely by writes, or vice-versa, because the protocol disallows
the same data item from being read and written concurrently to ensure a given
isolation property
(e.g., Google's Spanner strictly-serialisable transactions \cite{rep:pan:1693}).

\subsection{Freshness}
\label{sec:freshness}
Another important metric is how recent is the data returned by a read.
Users prefer recent data \cite{twitter}; some strongly-consistent criteria
(e.g., Strict Serialisability) 
might require data to be the latest version; and in others
(e.g., Snapshot Isolation) serving recent data makes aborts less likely and
hence improves overall throughput \cite{rep:syn:sh156,gmu}.
Storing only the most recent version of a data item enables
update-in-place and avoids the operational costs of managing multiple
versions.

However, many protocols require maintaining multiple
versions of a data item to read consistently.
Serving an old item state may be faster than waiting for the newest one to
become available; indeed, it would be easy for reads to be both fast and
consistent, by always returning the initial state.

\emph{Freshness} is a qualitative measure of whether snapshots include
recent updates or not.
The most aggressive is \textbf{Latest Freshness}, which guarantees that, 
for every data item that a server stores, 
it returns the most recent version that it has committed so far.
Intuitively, systems like Espresso and Tao \cite{tao,linkedin-espresso},
which do not make snapshot guarantees, can read with minimal
delay under Latest Freshness.
 
The most conservative is \textbf{Stable Freshness}, under which, 
for any read request in a given transaction, the server returns data taken 
from a snapshot that is \emph{stable} for that transaction, 
i.e., all versions included in the snapshot are known to be committed by 
the time the transaction starts.

The intermediate \textbf{Concurrent Freshness} does not necessarily
return the latest version, but allows a transaction to 
read updates committed after its starting point, i.e., 
not stable. For instance, a transaction $T_R$ can read updates
from a committed transaction $T_U$ that ran concurrently with $T_R$.
COPS is a system that exhibits
Concurrent Freshness \cite{rep:syn:1662}.

\subsection{Optimal reads}
\label{sec-optimal-reads}
We say a protocol provides optimal reads if it ensures both Minimal Delay and Latest Freshness.
An optimal-read protocol is one that supports parallel reads, and where a
server is always able to reply to a read request immediately, in a
single round trip, with the latest committed version that it stores.

%



\section{The three-way trade-off}
\label{sec-tradeoff}
In this section, we study the three-way trade-off 
between transactional reads semantics, delay and freshness.
In summary, our analysis concludes the following:

\begin{compactenum}[\it (i)]
\item
  {\it Impossibility of optimal order-preserving reads.}
  Ensuring optimal reads is not possible under
  Order-Preserving or Atomic Visibility (Section~\ref{sec-theimpo}).
\item
  {\it Order-Preserving Visibility with Minimal Delay and Concurrent Freshness.}
  Order-Preserving Visibility can ensure Concurrent Freshness at
  Minimal Delay (Section~\ref{sec-forward}).
\item
  {\it Atomic Visibility with Minimal Delay forces Stable Freshness.}
  To ensure Minimal Delay, Atomic Visibility forces transactions to
  read from a \emph{stable snapshot}, i.e., a snapshot consisting of updates known to
  have committed in the past (Section~\ref{sec-past}).
\item
  {\it Consistent reads with Latest Freshness.}
  To guarantee reading the freshest data, Order-Preserving and Atomic
  Visibilities require reads and updates mutually exclusive
  (Section~\ref{sec-latest-fresh-imp}).
  
\end{compactenum}

\subsection{Notation and Definitions}
\label{sec-notation}

\medskip\noindent\textbf{Notation.}
A committed update transaction creates a new version of
the data items it updates.
For some data item (or object) $x \in \mathcal{X}$, where $\mathcal{X}$
is the universe of object identifiers, we denote a version $x_v \in V$, 
where $V$ denotes the universe of versions.
We assume an initial state $\bot$ consisting of a initial version $x_{\bot}$
for every $x\in\mathcal{X}$.
If versions follow a partial or total order $O=(V, \prec)$,
we say a version $x_i$ is more up-to-date (or fresher) than
a version $y_j$ when $y_j \prec x_i$.

The database is partitioned, i.e., its state is divided into $P \ge 1$
disjoint subsets, where all the versions of a given object belong to
the same partition.
Throughout the text, we use the terms partition, server and storage server
interchangeably.


\medskip\noindent\textbf{Definitions.}
We define the three types of snapshots introduced
 in subsection \ref{sec-apguarantees} formally:
\begin{definition}[Committed snapshot]
  A committed snapshot $S$ is any subset of $V$ that includes exactly
  one version of every object $x\in\mathcal{X}$.
  $\mathcal{S}$ denotes the set of all committed snapshots.
\end{definition}

\begin{definition}[Order-Preserving Snapshot]
  \label{def-ordcon}
  Given a partial or total order of versions $O=(V, \prec)$, a committed snapshot $S_O \in
  \mathcal{S}$ preserves $O$ if $\forall x_i, y_j\in S_O$,
  $\nexists x_k\in V$ such that $x_i \prec x_k \prec y_j$.
  Intuitively, there is no gap in the order of versions visible in an order-preserving
  snapshot.
  We denote $\mathcal{S}_O \subseteq \mathcal{S}$ the set of committed
  snapshots preserving order $O$.
\end{definition} 

\begin{definition}[Atomic Snapshot]
  \label{def-acsnap}
  Given an order $O$, an order-preserving snapshot $S_A\in\mathcal{S}_O$ is
  atomic if $\forall x_i, y_j\in V$ such that $x_i, y_j$ were written by
  the same transaction, if $x_i, y_k \in S_A$ then $y_k \nprec y_j$, i.e.,
  it disallows ``broken reads''.
  We denote $\mathcal{S_{A}}$ the set of atomic snapshots for order
  $O$, $\mathcal{S_{A}} \subseteq \mathcal{S}_O$.
\end{definition}

\begin{definition}[Snapshot guarantee]
  Given some order $O$, we say that a read protocol guarantees Committed
  (, Order-Preserving or Atomic) Visibility if it guarantees
  that every transaction reads from a Committed (, Order-Preserving, or Atomic, respectively)
  snapshot.
\end{definition}

\begin{figure}[tp]
  \centering
  \includegraphics[width=.65\columnwidth]{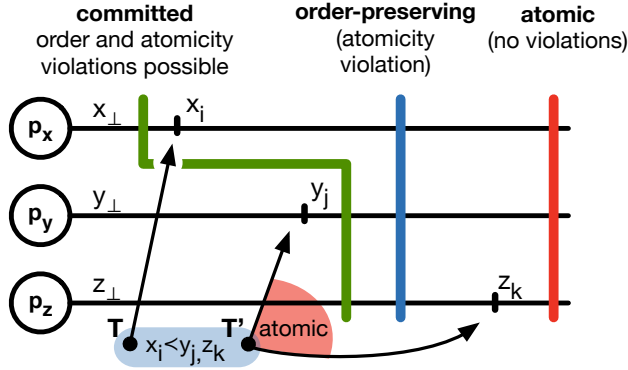}
  \caption{The three snapshot guarantees}
  \label{fig:snapshots}
\end{figure}

We illustrate the three types of snapshots in
Figure~\ref{fig:snapshots}.
The figure shows a system consisting of three partitions,
$p_{x}$, $p_{y}$, and $p_{z}$, each storing a single object
$x$, $y$, and $z$, in an initial state $x_{\bot}$, $y_{\bot}$, $z_{\bot}$, respectively. 
Two transactions $T$ and $T'$ have committed 
updates in order $x_{\bot}, y_{\bot}, z_{\bot} \prec x_i\prec y_j,z_k$.
$T$ updates only partition $p_{x}$, whereas $T'$ updates $p_{y}$ and
$p_{z}$ atomically.
The figure highlights three possible snapshots.
Under Atomic Visibility, only the atomic snapshot is admissible, precluding both
order violation (both $T$ and $T'$'s updates are included) and read skew
(as both $y_j$ and $z_k$ are included).
Under Order-Preserving Visibility, the atomic and the
order-preserving snapshots are both admissible.
The latter precludes order violations, but not read skews (e.g.,
the snapshot includes $y_j$ from transaction $T'$, and not $z_k$).
Finally, under Committed Visibility, all three depicted snapshots are admissible
because order violations and read skews are allowed.
The snapshot at the left of the picture exhibits two anomalies:
an order violation, and a read skew. 
The order violation occurs by reading $x_{\bot}$ and $y_j$, as 
$x_{\bot} \prec x_i \prec y_j$, and $x_i$ is not read.
The read skew occurs by reading $z_{\bot}$ and $y_j$, as 
$y_j$ was created atomically with $z_k$, which is not read.

\subsection{Impossibility of optimal order-preserving reads}
\label{sec-theimpo}
\begin{proposition}
  \label{prop-impossible}
  Order-Preserving (or Atomic) reads cannot
  ensure optimal (delay and freshness).
\end{proposition}

\noindent{\it Proof.} We prove this proposition by contradiction.
Assume that there exists a read-optimal protocol that guarantees
Order-Preserving (or Atomic) Visibility, w.r.t.~order $O=(V,\prec)$.
Consider the execution in Figure~\ref{fig:impossibility} where,
initially, partition $p_x$ stores $x_{\bot}$ and $p_y$ stores $y_{\bot}$.
Two transactions $T_u$ and $T_{u'}$ write $x_k$ at $p_x$
and $y_j$ at $p_y$ respectively, 
establishing the following order: $x_{\bot}, y_{\bot} \prec x_k \prec y_j$.
For instance, under causal order, this can result from an 
execution where a transaction reads $x_{\bot}$, and updates $x$, creating
$x_k$, and later another transaction reads $x_k$
and updates $y$, creating $y_j$.
A $T_r$, running concurrently with $T_u$ and $T_{u'}$,
reads objects $x$ and $y$ in parallel from 
$p_x$ and $p_y$.
$T_r$ reaches $p_x$
before the creation of $x_k$, and $p_y$ after the creation of $y_j$.
To satisfy read optimality, partitions must reply immediately 
with the latest version they store, namely $x_{\bot}$ and $y_j$, 
observing an order violation%
\footnote{
	A similar situation occurs with the execution of Figure
	\ref{fig:impossibility-atomic}, where reading $x_{\bot}$ and
	$y_j$ results in a read skew.
}
.
Contradiction. $\blacksquare$

\subsection{Freshness compatible with Minimal Delay}
In this subsection we explore which are the
maximum freshness degrees achievable for each snapshot 
guarantee, under the requirement of Minimal Delay.

\subsubsection{Optimal reads under Committed Visibility}

\begin{proposition}
	\label{prop-ar}
A read protocol that guarantees Committed Visibility can be
optimal.
\end{proposition}

\noindent{\it Proof.} Committed Visibility imposes no restrictions 
to the committed versions a transaction can read.
Therefore, to serve a request under this model, a partition 
can reply immediately with the
latest object version it stores. $\blacksquare$

\noindent
\subsubsection{Order-Preserving Visibility and Concurrent Freshness}
\label{sec-forward}
\begin{proposition}
	\label{prop-cr}
Order-preserving minimal-delay reads
can ensure Concurrent Freshness.
\end{proposition}

We prove this proposition by sketching a read protocol with such
characteristics, followed by a
correctness proof.
In Section \ref{sec-protocols}, we present a protocol with these
characteristics.

\begin{figure}
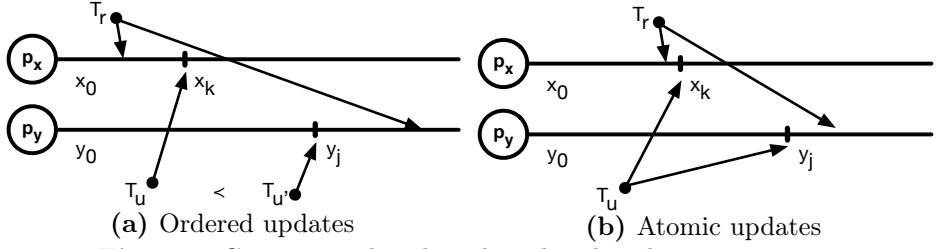

  \centering
      \begin{subfigure}{0.48\columnwidth}
		  \centering
  	  	\includegraphics[width=\linewidth]{figures/tradeoff-pic.pdf}
  	  	\caption{Ordered updates}
  	  	\label{fig:impossibility}
  	  \end{subfigure}
      \begin{subfigure}{0.48\columnwidth}
		  \centering
  \includegraphics[width=\linewidth]{figures/tradeoff-pic2.pdf}
  \caption{Atomic updates}
  \label{fig:impossibility-atomic}
  	  \end{subfigure}
  \caption{Concurrent distributed read and update transactions}  
\end{figure}

Consider a protocol that orders its updates 
following some order $O=(V,\prec)$, and where reads
preserve $O$.
When a read transaction starts, the protocol assigns it
an $O$-preserving \textit{stable} snapshot $S_{O}$ 
(subsection \ref{sec:freshness}).
Read requests are sent to their corresponding partition in parallel.
A partition can reply immediately with the version in $S_{O}$
or with a more up-to-date version that is
\emph{compatible} with $S_{O}$.
An object version $y_j$ is {compatible} with a given
order-preserving snapshot $S_O$ if replacing version 
$y_s \in S_O$ by $y_j$ results in an order-preserving snapshot.
Formally:
\begin{definition}[Compatible version]
  \label{def-compatible}
  Given an order $O=(V, \prec)$, a version $y_j \in V$
  and an order-preserving snapshot $S_O$,
  an object version $y_j \notin S_O$ is compatible with $S_O$, if $\forall
  x_i \in S_O$, $\nexists x_k \in V$ such that $x_i \prec x_k \prec y_j$.
\end{definition}

\begin{lemma}
\label{lemma:compatible}
Given an order-preserving snapshot $S_O$,
replacing any number of versions $x_o \in S_O$ by $x_i \notin S_O$,
such that $x_i$ is compatible with $S_O$,
results in an order-preserving snapshot $S_{O'}$.
\end{lemma}

\noindent{\it Proof.} 
Assume by contradiction that the resulting snapshot $S_{O'}$ 
is not order-preserving
w.r.t.~order $O=(V,\prec)$. 
According to Definition~\ref{def-ordcon}, this implies
that $\exists x_i, y_j \in S_{O'}, x_k \in V : x_i \prec x_k \prec y_j$. 
Since $S_O$ is order-preserving,
if the versions returned by read partitions
were those in $S_O$, i.e., $x_o$ and $y_o$,
no inconsistency could have been created.
Now consider the case where only one compatible version with $S_O$, 
e.g., $y_j$, is more up-to-date than $y_o \in S_O$
($y_o \prec y_j$).
By Definition~\ref{def-compatible},
$\nexists x_k \in V :x_o \prec x_k \prec y_j$. 
Finally, assume that both $x_i, y_j \notin S_O$
are more up-to-date compatible versions of 
objects $x$ and $y$.
As they are compatible with $S_O$,
by Definition~\ref{def-compatible}, 
\emph{(i)}$\nexists x_k: x_o \prec x_k \prec y_j$ and
$\nexists y_l: y_o \prec y_l \prec x_i$.  
Moreover, we know that \emph{(ii)}$x_o \prec x_i$ 
and $y_o \prec y_j$.
\emph{(i)} and \emph{(ii)} 
imply $y_j \nprec x_i$ and $x_i \nprec y_j$.
Therefore, there cannot exist 
$x_k : x_i \prec x_k \prec y_j$.
Contradiction. $\blacksquare$

\begin{lemma}
\label{lemma:correctness}
The above
protocol guarantees Order-Preserving Visibility.
\end{lemma}

\noindent{\it Proof.} This follows directly from
Lemma~\ref{lemma:compatible}. $\blacksquare$

\begin{lemma}
\label{lemma:ff}
The above protocol allows Concurrent Freshness.
\end{lemma}

\noindent{\it Proof.} We prove this lemma by describing a sample execution.
Assume transaction $T_r$ starts with stable snapshot $S=\{x_o, y_o\}$.
$T_r$ sends a read request for objects $x$ and $y$ to partitions $p_x$
and $p_y$ respectively.
Concurrently, update transactions create versions $x_u$ and $y_v$
establishing the following order between them: $x_o, y_o
\prec x_u \prec y_v$.
$T_r$'s request arrives to $p_x$ after $x_u$ and $y_v$ are committed.
By Definition~\ref{def-compatible}, $p_x$ can reply with $x_u$, a more
up-to-date version.
However, $p_y$ can only reply with $y_o$, as $y_v$ is not compatible
with $S$ ($\exists x_u\in V : x_o (\in S) \prec x_u \prec
y_v$).
As $x_u$ is committed by an update transaction concurrent to $T_r$, this
execution exhibits Concurrent Freshness.
$\blacksquare$

\begin{lemma}
\label{lemma:minimaldelay}
The above
protocol guarantees minimal delays
\end{lemma}

\noindent{\it Proof.} The protocol reads 
versions in parallel.
In the absence of fresher committed updates than those
in $S_O$, a partition can reply immediately with versions belonging to $S_O$,
which is stable and, therefore, already committed.
In the presence of fresher and compatible committed updates, a partition can 
reply to a request with those, immediately. 
$\blacksquare$

\medbreak
\noindent{\it Proof of Proposition~\ref{prop-cr}}. This
follows directly from
Lemmas~\ref{lemma:compatible}, \ref{lemma:minimaldelay}, \ref{lemma:correctness} and \ref{lemma:ff}.
$\blacksquare$

\subsubsection{Stable Freshness under Minimal-delay Atomic Visibility}
\label{sec-past}
\begin{proposition}
  \label{prop-av}
  A minimal-delay read protocol that guarantees Atomic Visibility
  requires Stable Freshness.
\end{proposition}

The intuition is that, due to the minimal-delay requirement, 
a partition receiving a read request from a transaction $T_r$
cannot know whether other partitions accessed by $T_r$ 
are returning updates of a concurrent update transaction $T_u$ or not, 
which forces it to read from a stable snapshot to avoid Read Skews.

\medbreak
\noindent{\it Proof.} Assume by contradiction that there exists a
minimal-delay protocol that guarantees Atomic Visibility and
allows a transaction to read updates committed by other 
concurrent transactions (Concurrent Freshness).
Consider the example execution in Figure \ref{fig:impossibility-atomic}.
A transaction $T_r$ sends parallel requests to read 
objects $x$ and $y$ from partitions $p_x$ and
$p_y$ respectively.
A concurrent transaction $T_u$ commits versions $x_k$ and $y_j$.
Assume that $T_r$'s request reaches $p_y$ after $T_u$ commits.
By Definition~\ref{def-acsnap}, $p_y$ can return $y_j$ only if it is
certain that $T_r$ will read $x_k$ from $p_x$. 
Due to the
minimal-delay requirements, $p_x$ does not have access to such
information, since reads can be executed in parallel and no extra communication among
partitions is allowed. 
Given that $T_r$ can reach $p_x$ before $T_u$ commits
$x_k$, $p_y$ cannot risk returning $y_j$, and must ignore $T_u$. 
Therefore, a partition can only return a version of
an update transaction $T_u$ if it knows $T_u$ had committed at 
all its updated partitions by the time $T_r$ sent its read requests.
This implies that $T_r$ has to read from stable snapshot,
which contradicts our assumptions.
$\blacksquare$


\subsection{What is possible under Latest Freshness?}
\label{sec-latest-fresh-imp}

\begin{proposition}
\label{prop-latest-freshness}
Order-Preserving (and Atomic) Visibility require mutually-exclusive
reads and updates to guarantee Latest Freshness.
\end{proposition}

\begin{lemma}
  \label{lemma:bounded}
It is not possible to guarantee
Order-Preserving or Atomic Visibility with Latest Freshness under
Bounded Delay.
  \end{lemma}

\noindent{\it Proof. } 
Consider again the sample execution of Figure~\ref{fig:impossibility}
(where $x_{\bot}, y_{\bot} \prec x_k \prec y_j$).
To ensure Latest Freshness, partitions must reply to read requests
with the latest committed version they store.
If $p_x$ returned $x_{\bot}$ and $p_y$ returned $y_j$, $T_r$ 
would observe an inconsistent result (by missing $x_k$).
The protocol could retry reading from $p_x$ to read $x_k$, 
thus ensuring reading a version compatible with $y_j$.
If such request arrived to $p_x$ before $T_u$ created $x_k$, 
$p_x$ could block until $x_k$ was applied to read it.
During the blocking period, a concurrent update transaction may have
written a new version $x_m$ such that $y_j \prec y_w \prec x_m$.
To satisfy Latest Freshness, $p_x$ would be forced to reply with $x_m$,
inconsistent with the version read from $p_y$: $y_j$.
If updates are not stopped, this situation can repeat itself indefinitely, 
making reading with Bounded Delay impossible. $\blacksquare$


\begin{lemma}
  \label{lemma:possiblemutual}
A read protocol can ensure Order-Preserving (and Atomic) Visibility
and Latest Freshness by
enforcing mutually-exclusive reads and updates.
\end{lemma}

\noindent{\it Proof.} 
We prove this is possible by following the proof of Lemma \ref{lemma:bounded}.
In the execution of Figure \ref{fig:impossibility}, $T_r$ can retry 
indefinitely reading the latest versions of $x$ and $y$ until the results
belong to an order-preserving snapshot. 
The equivalent holds for building an atomic snapshot under
the execution of Figure \ref{fig:impossibility-atomic}. $\blacksquare$


\medbreak
\noindent{\it Proof of Proposition~\ref{prop-latest-freshness}}. This
follows directly from Proposition~\ref{prop-impossible}, and Lemmas~\ref{lemma:possiblemutual} and~\ref{lemma:bounded}.

\subsection{Isolated reads with Bounded Delay and Concurrent Freshness.}
\label{av-bounded-forward}

\begin{lemma}
  \label{lemma:possibleconcurrent}
A read protocol can ensure Order-Preserving (and Atomic) Visibility
and Concurrent Freshness under Bounded Delay.
\end{lemma}

\noindent{\it Proof. } 
Consider again the execution in Figure~\ref{fig:impossibility} where
read transaction $T_r$ executes concurrently with update transactions
$T_u$ and $T_{u'}$.
If $p_x$ returns $x_{\bot}$ and $p_y$ returns $y_j$, it is possible
to issue a second round to force $p_x$ to return $x_k$, 
which would ensure Order-Preserving Visibility and Concurrent Freshness.
The same holds for ensuring Atomic Visibility in the example execution
in Figure \ref{fig:impossibility-atomic}. $\blacksquare$

Many existing systems exhibit these properties (see Section \ref{sec:systems}). 



\section{Minimal-delay protocol design}
\label{sec-protocols}

In this section, we apply our trade-off analysis to the design of protocols.
We design three novel minimal-delay protocols, called CV, OP and AV.
Our protocols are a modification of Cure \cite{rep:pro:sh182}.
\SYSTEM{} ensures Transactional Causal
Consistency (TCC), i.e., Causal Consistency 
(where updates follow causal order) and snapshot reads
ensure Atomic Visibility.
It exhibits Bounded Delays and Concurrent Freshness.
We use the insights of the analysis: to remove the delays
of \SYSTEM{} one must degrade either read semantics or freshness. 
AV provides the same isolation guarantees as the base protocol.
It achieves Minimal Delay by degrading freshness to stable.
OP is the first protocol to provide causally-consistent snapshot reads 
(Order-Preserving Visibility) and atomic updates.
It exhibits Concurrent Freshness.
CV ensures Read Committed Isolation, i.e., Committed Visibility
and atomic updates. It achieves Latest Freshness.

\subsection{Base protocol and changes}
\label{base-protocol}
Cure associates a snapshot timestamp 
to a transaction when it starts.
When committing updates, a transaction creates object versions which are stored with
a commit timestamp.
Timestamps are vector clocks sized with the number of sites.
Each position in a vector is assigned by the protocol using the physical clocks of servers.

\noindent \textbf{Updates.} The protocol can be characterised as \emph{Deferred Update
Replication} (DUR), i.e., a transaction buffers updates while executing and
sends them to storage servers when committing \cite{dur}.
A 2PC ensures that the updates of a transaction are applied in 
an all-or-nothing fashion, and that commit
timestamps respect causal order. I.e., that a transaction's commit timestamp
is bigger than that of the object versions the transaction read.

\noindent \textbf{Reads.}
When reading, a transaction sends a read request to a partition including
its snapshot timestamp. 
A partition receiving a request must reply with the version 
with the maximum commit timestamp smaller than the snapshot timestamp of the request.  
In Cure, the snapshot timestamp associated to transactions
is not necessarily stable. 
A partition that receives a read request must delay a response when
(i) an instance of 2PC is committing a transaction wich needs to be included in the snapshot, and/or
(ii) the clock at the partition is behind the snapshot timestamp.

\noindent\emph{Changes.} Our protocols remove the blokcing situations of \SYSTEM\ by applying
the following modifications:
\begin{itemize}[leftmargin=*]
\item AV degrades freshness.
Specifically, it ensures that the snapshot assigned 
to a transaction is stable. 

\item OP degrades visibility to Order-Preserving.
Given an initial stable snapshot, 
a partition is allowed to return a more up-to-date compatible version
(Definition \ref{def-compatible}).

\item CV degrades read guarantees to Committed Visibility,
allowing a partition to return the latest committed version.
\end{itemize}

To minimise delays of update transactions,
all protocols implement a low-latency 2PC
protocol, called presumed commit \cite{pres-commit}.
Under this optimisation, 
a transaction is considered to be committed
after a successful prepare phase, i.e., 
after every involved partition has
persisted a prepare record including its updates.
A client can, then, receive a response after a single
round-trip of communication with updated partitions.

\subsection{Protocols}
\noindent\textbf{Setting and notation.}
All designs consider $M$ fully-replicated sites.
Every site partitions data into $N$ partitions.
All sites follow the same partitioning scheme,
i.e., for each partition $p_n^m, n=1...N$ at a given site $m$,
there exists a partition $p_n^{k}$ storing the same objects at
every other site $k: k=1...M, k \neq m$.
Updates are replicated across sites asynchronously. 
We refer to partitions storing the same set of objects at different sites as
\emph{sibling} partitions.
Each transaction executes entirely within a site.

\smallskip\noindent\textbf{API.}
The data access APIs are
multi reads and multi updates. 
A client can group any number of 
read or update operations respectively, 
and execute them against the
storage servers in parallel.
A transaction comprises any number of such multi-operations.

All algorithms share a general skeleton. 
In what follows, we look at their points in common.
In Section \ref{sec-alg-particularities}, we address each algorithm's
particularities. 
There are two types of processes involved in a transaction's 
execution: a \emph{transaction coordinator (TC)}, and the
\emph{partition} servers.

\subsubsection{Transaction Coordinator Algorithm}

\begin{algorithm}[tp] 
	\footnotesize
  \begin{algorithmic}[1]		  
    \Function{read\_objects}{keys} \label{tcread}
	   \State \textbf{If} (!initiated) \Call{init}{ }
	   \State result = $\emptyset$
		\State read\_partitions$_T$=\Call{GET\_PARTS}{keys}\label{tcgetpar}
		\ForAll{$\langle$p, keys$_p$ $\rangle$ $\in$ partitions$_T$}
			\State send $\langle$\textbf{read}, keys$_p$, $ss_T$$\rangle$ to p
			\label{tcsendread}
		\EndFor
		\ForAll{$\langle$p, keys$_p$ $\rangle$ $\in$ partitions$_T$} \label{tcreceivereads}
			\State receive $\langle$partition\_result$\rangle$ from p
			\State result = result $\cup$  \{partition\_result\} 
		\EndFor
		\State \textbf{If} (protocol == OP) \label{tcupdatedep}
		\Indent
			\State commit\_vc=\Call{max$_v$}{v.commit\_vc $\in$ result} \label{tcupdatemaxver}
			\State $dep_T$=\Call{max$_v$}{$dep_T$, commit\_vc}\label{tcupdatedepend}
		\EndIndent
		\State \Return result.values
	\EndFunction
	
		\smallskip
	
    \Function{update\_objects}{$[\langle$key, update$\rangle]$}
   \State \textbf{If} (!initiated) \textbf{then} \Call{init}{ }
	\State $WS_T$ = $WS_T$ $\cup$ \{$[\langle$key, update$\rangle]$\}\label{tcupdate}
	\State \Return ok
	\EndFunction

		\smallskip
	
	\Function{commit}{ }\label{tccommit}
		\State $ct_T$=$dep_T$[n] $+ 1$ \label{algtccausalct}
		\State updated\_partitions$_T$ = \Call{GET\_PARTS}{$WS_T$.keys}
		\ForAll{$\langle$p, updates$_p$ $\rangle$ $\in$ updated\_partitions$_T$}\label{tc2pcstart}
			\State send $\langle$\textbf{prepare}, updates$_p$, $dep_T, ct_T$$\rangle$ to p\label{tcpreparesend}
		\EndFor
		\ForAll{$\langle$p, keys$_p$ $\rangle$ $\in$ updated\_partitions$_T$}
			\State receive $\langle$\textbf{prepared}, time$_p \rangle$ from p
			\State \textbf{If} (protocol $\neq$ CV) \textbf{then} $ct_T$=\Call{max}{$ct_T$, time$_p$}\label{tcmaxprepare}
		\EndFor
		\State \textbf{send} ok to client
		\ForAll{$\langle$p, updates$_p$ $\rangle$ $\in$ updated\_partitions$_T$}
			\State send $\langle$\textbf{commit}, $ct_T$$\rangle$ to p \label{tc2pcend}
		\EndFor
		\State \Call{terminate}{ }
	\EndFunction
	
		\smallskip

  	\Function{init}{ } \label{tcinit}
  	\State $WS_T$=$\emptyset$
	\State \textbf{If} (protocol == OP or AV)
	\Indent
		\State $ct_T$ = $\bot$
		\State $dep_T$ = $ss_T$ = \Call{get\_stable\_vector}{ }\label{tcinitdep}
		\State initiated = $true$ \label{tcinitend}
	\EndIndent 
  	\EndFunction
	
  \end{algorithmic}
  \caption{Transaction Coordinator tc at site n}
  \label{alg-coord}
\end{algorithm}

The server that receives 
a new transaction starts a new TC process.
The TC lives throughout the execution of the transaction
and terminates once the transaction commits or aborts.


\smallskip\noindent\textbf{Init.}
A TC initialises (Algorithm~\ref{alg-coord}, Lines~\ref{tcinit}-\ref{tcinitend}) 
on a client's first read or update request. 
It initialises the write set for the transaction $WS_T$, the associated
snapshot $ss_T$, the transaction's commit time $ct_T$, and the dependency vector clock $dep_T$
(used for creating a causal order of updates).


\smallskip\noindent\textbf{Reads.}
When receiving a read request for a list of
object keys (Line~\ref{tcread}),
the TC splits them by partition by calling the
$GET\_PARTS$ function (Line~\ref{tcgetpar})
and sends a read request to each partition in parallel (Line~\ref{tcsendread}).
Once it receives all responses (Line~\ref{tcreceivereads}),
it returns the read values to the client.

\smallskip\noindent\textbf{Updates and commit.}
When a client submits an update, the transaction coordinator
buffers it (Line~\ref{tcupdate}) in its write set.
If the transaction updates the same data item multiple times, 
they are applied (by the commit protocol) in the order they are submitted.
This is useful when updating data structures such as counters, sets, lists, etc.
When the client calls commit (Line~\ref{tccommit}), 
if the transaction updated multiple partitions, the TC starts
an instance of the 2PC protocol among the updated partitions
(Lines~\ref{tc2pcstart}-\ref{tc2pcend}).
In the prepare phase, the TC sends, in parallel, 
a prepare message containing partition's updates
to each updated partition (Line~\ref{tcpreparesend}).
The participants reply with a prepare time, 
a proposed commit timestamp for the transaction. 
Once it has received the response from every participant, the TC
\begin{inparaenum}[i)]
	\item
	computes the commit time of the transaction as the maximum proposed
	prepare time,
	\item replies to the client confirming that the transaction has committed, and
	\item sends the commit instruction, including the commit timestamp,
 to all participants.
\end{inparaenum}
If the transaction updated a single partition, we collapse
the prepare and commit messages.
This optimisation is not depicted in the pseudocode.
An abort will occur only if requested by the client, or in 
case of a failure. 
The abort path discards the transaction updates.
The protocol, identical to that of presumed commit \cite{pres-commit},
is not depicted 
in the pseudocode. 

\subsubsection{Partition Servers Algorithm}

\begin{algorithm}[tp]
	\footnotesize
  \begin{algorithmic}[1]
			  \State \textbf{upon} receive $\langle$\textbf{read}, keys,$ss_T$$\rangle$ from tc
			  \Indent
			  	\State result=$\emptyset$
				\ForAll{k $\in$ keys}
					\State v= newest k$_i$ $\in$ ver[k] : \Call{cond}{k$_i$, $ss_T$}
					\State result = result $\cup$ \{v\}
		\EndFor
		\State \textbf{send} $\langle$result$\rangle$ to tc
			
		\EndIndent

		\smallskip

		\State \textbf{upon} receive $\langle$\textbf{prepare}, upd, $dep_T, ct_T$$\rangle$ from tc \label{algpprepare}
		\Indent
			\State \textbf{If} (protocol $\neq$ CV)
			\Indent
				\State time =  {\small MAX}(\Call{read\_clock}{ }, $ct_T$) \label{preptime}
				\State prepared = prepared $\cup$ $\langle$tc, $upd$, $dep_T$, time$\rangle$
			\EndIndent \label{algplogphy}
			\State \textbf{Else} prepared = prepared $\cup$ $\langle$tc, $upd$$\rangle$
			\State \textbf{send} $\langle$\textbf{prepared}, time$\rangle$ to tc
		\EndIndent

		\smallskip
		
		\State \textbf{upon} receive $\langle$\textbf{commit}, $ct_T$$\rangle$ from tc\label{algpcommit}
		\Indent
			\State prepared = prepared $\setminus $$\langle$tc, $upd$, $dep_T$, time$\rangle$
			\State \Call{update\_versions}{$\langle$$upd$, $dep_T$, $ct_T$, n$\rangle$} 
			\State to\_send=to\_send $\cup$ \{$\langle$upd, $dep_T$, ct$\rangle$\} \label{algpsubmit}
		\EndIndent
		
		\smallskip
		
		\Function{cond}{k$_i$, $dep_T$}\label{algpcondbegin}
			\State \textbf{If} (protocol==CV) \textbf{then} \Return true \label{algpreadar}
			\State \textbf{If} (protocol==OP) \textbf{then} \Return k$_i$.dep $\leq$ $dep_T$ \label{algpreadcr}
			\State \textbf{Else} \Return  k$_i$.cv $\leq$ $dep_T$\label{algpcondend}\label{algpreadav}
		\EndFunction
		
		\smallskip
		
		\Function{update\_versions}{upd, $dep_T$, ct, n}\label{algupdbegin}
			\ForAll{$\langle$k,val$\rangle$ $in$ upd}
				\State \textbf{If} (protocol = CV) \textbf{then} ver[k] = \{val\}
				\State \textbf{Else} ver[k] = ver[k]$\cup$ \{$\langle$val, $dep_T$, ct, n$\rangle$\}
			\EndFor 
		\EndFunction
  \end{algorithmic}
  \caption{Partition $m$ at site $n$ $p_m^n$}
  \label{alg-partition}
\end{algorithm}

A partition server stores versions of the objects in its partition
and replies to requests from transaction coordinators that access those objects.

\smallskip\noindent\textbf{Reads.}
When a partition receives a read request from a TC,
it replies with the most up-to-date version of each requested object
that satisfies the algorithms' visibility criteria
(Lines~\ref{algpcondbegin}-\ref{algpcondend}).

\smallskip\noindent\textbf{Updates and commit.}
A partition server participates in instances of 2PC coordinated
by some TC.
In the first phase, 
the partition server receives a prepare message 
containing updates (Line~\ref{algpprepare}).
It persists those updates to stable storage 
(not depicted in the pseudocode),
and replies with a \emph{prepared}
message; a positive vote to commit the transaction.
At this stage, these updates are persisted but not 
accessible by other
transactions, as the transaction has not yet committed.
Later, if the partition server
receives a commit message (Line~\ref{algpcommit}),
it makes updates visible to future readers.

\smallskip\noindent\textbf{Update propagation.}
Under replication, a partition server propagates committed 
updates asynchronously to sibling partitions at remote sites.
When a partition receives a remote transaction's updates,
it applies them locally and makes them available to future read
operations.
CV makes updates visible as they arrive. 
The stabilisation protocol run by AV and OP ensures a remote
update is made visible to readers respecting causal consistency,
as explained in the next section. 

\subsection{Correctness}
\label{sec-alg-particularities}
We discuss here the differences between the three
protocols and focus on their correctness.

\medbreak
\noindent{\bf Consistent Reads.} 
OP and AV provide across-object isolation. 
When a TC executes the first read (or update), it assigns
the transaction a snapshot timestamp $ss_T$,
which is used as a pivot to compute versions consistent
with the target isolation level.
$ss_T$ is assigned the partition's stable vector $SV_n^m$, 
which denotes the latest stable snapshot known
by the partition where the TC runs 
(Line~\ref{tcinitdep}). In Section \ref{sec:stabilisation}, we describe how each
partition maintains this vector.

A TC includes $ss_T$ in each read request it sends
to partition servers (Line~\ref{tcsendread}).
When a server receives a request, it responds with
the most up-to-date version
that complies with the requested snapshot, according to a
$COND$ function parametrised by protocol:

\begin{itemize}[leftmargin=*]
\item Under CV, $COND$ returns the latest committed version.
  
\item Under AV, $COND$ ensures that versions do not violate 
an atomic snapshot that preserves causal order.
$COND$ returns
the newest version with a commit vector smaller than $ss_T$
(Alg.~\ref{alg-partition}, Line~\ref{algpreadav}).
It ensures causal consistency, 
since the snapshot is stable, i.e.,
all updates with $cv \leq ss_T$ have been applied.
It satisfies Minimal Delay, as partitions can reply immediately.
It also guarantees Atomic Visibility (the absence of fractured reads)
because all updates of a given transaction commit with the 
same commit timestamp (as explained later in this section).

\item Under OP, $COND$ ensures that versions belong to
a causal-order-preserving snapshot.
A partition server returns either the version belonging to the
stable snapshot $ss_T$, as above, 
or with a more up-to-date version compatible with $ss_T$, if available
(see Definition~\ref{def-compatible}).
A version is compatible if 
its associated dependency vector is not larger
than $ss_T$
(Alg.~\ref{alg-partition}, Line~\ref{algpreadcr}).



\end{itemize}

AV and OP provide causal consistency, which ensures all session
guarantees \cite{rep:syn:1481}. 
We explain how these are ensured in Appendix \ref{app:sesgar}.

\medbreak
\noindent{\bf Causal order of updates.}
OP and AV ensure updates are causally ordered.
A transaction creates an object version with
a commit timestamp, and a
dependency vector $dep$.
$dep$ indicates a version is ordered
causally after all versions with commit vector $cv \leq dep$.
A commit vector $cv$ is created by replacing,
in $dep$, the entry of the site where the version was committed by its
commit timestamp $ct$.
To establish a correct causal order, these algorithms must 
ensure a version's $cv$ is larger than its $dep$.
The transaction coordinator ensures this by picking 
a $ct$ which is larger than the transaction dependencies
(in Alg.~\ref{alg-coord}, Line~\ref{algtccausalct}).
To ensure that $dep$ is larger than the $cv$ of its 
predecessors:
\begin{itemize}[leftmargin=*]  
\item Under OP, a transaction may read a version
with $cv$ larger than $ss_T$.
After receiving a read response, the TC
updates the transaction's dependency vector $dep_T$
to the maximum $cv$ of a version read
(Alg.~\ref{alg-coord}, Line~\ref{tcupdatedepend}).

\item AV's read algorithm ensures that a transaction will never
read a version with $cv$ larger than $ss_T$.
Therefore, it suffices to assign $ss_T$ to the transaction's
dependency vector $dep_T$ (Alg.~\ref{alg-coord},
Line~\ref{tcinitdep}). 
\end{itemize}

As Cure, OP and AV ensure that every update of a transaction
is assigned the same $cv$ by choosing a transaction's $ct$ 
as the maximum proposed time by an updated partition
(Alg.~\ref{alg-coord}, Line~\ref{tcmaxprepare}).
This is used by AV's read protocol to ensure Atomic Visibility.

\subsection{Stabilisation protocol}
\label{sec:stabilisation}

\begin{algorithm}[tp]
	\footnotesize
  \begin{algorithmic}[1]
		\State \textbf{periodically} \label{algpstabilise}
		\Indent
			\State \textbf{If} (prepared $\neq \emptyset$)
			\Indent
				\State stable$_n$ = \Call{min}{time $\in$ prepared} - 1 \label{algpminprepared} 
				
			\EndIndent

			\State \textbf{Else} stable$_n$ = \Call{read\_clock}{ }
			\State $vec_p$[n] = stable$_n$
			\State send $\langle$\textbf{stable}, $vec_p$$\rangle$ to p$_k^n$, $\forall k \in P, k\neq m$	\label{algpsendst}
			
			\State \textbf{If} (to\_send $\neq \emptyset$) 
			\Indent
				\ForAll{$\langle$upd$_p$, $dep$, ct$\rangle \in$ to\_send $: ct \leq stable_n$}	\label{algppropcausal}
				\State send $\langle$\textbf{updates}, upd$_p$, $dep$, ct$\rangle$ to p$_m^j$$j\neq n$	 \label{algppropagate}
				\EndFor
			\EndIndent
			\State \textbf{Else} send $\langle$\textbf{heartbeat}, stable$_n$$\rangle$ to p$_m^j$, $j\neq n$ \label{heartbeat}
			 
		\EndIndent
		
		\smallskip
		
		\State \textbf{upon} receive $\langle$\textbf{stable}, $vec_p$$\rangle$ from \textbf{all} p$_k^n$, $k\neq m$	
		\Indent
			\State $SV_m^n$=\Call{min$_v$}{$vec_p$}, \# $\forall p^j_m$\label{algpstablecompute}
		\EndIndent
		
		\smallskip
		\State \textbf{upon} receive $\langle$\textbf{updates},updates, $dep$, ct$\rangle$ from $p^j_m$ \label{algpreceiveremote}
		\Indent
			\State\Call{update\_versions}{updates, $dep$, ct, n}
			\State $vec_p$[j] = ct \label{algpremotestable}
			\State \# update known committed transactions from site j
		\EndIndent

		\smallskip
		\State \textbf{upon} receive $\langle$\textbf{heartbeat},stable$_j\rangle$ from $p^j_m$ \label{algpreceivehbeat}
		\Indent
			\State $vec_p$[j] = stable$_j$ \label{algpremotestableheart}
			\State \# update stable time from site j
		\EndIndent
		
  \end{algorithmic}
  \caption{Stabilisation for AV and OP at p$_m^n$}
  \label{alg-stabilisation}
\end{algorithm}

Under OP and AV, a transaction uses the knowledge of a stable
snapshot to read consistently and with Minimal Delay.
A stabilisation protocol among all partitions 
in the system computes stable snapshots at each site periodically.
Algorithm~\ref{alg-stabilisation} describes this protocol. It
includes the following steps:
\begin{enumerate}
\item When a partition server commits a transaction, it adds the transaction's
updates to {\small to\_send}, the list of updates to be sent to sibling
partitions at remote sites (Alg.~\ref{alg-partition}, Line~\ref{algpsubmit}). 
\item Periodically, updates in {\small to\_send} 
are propagated to sibling partitions (Line~\ref{algppropagate}). 
A partition sends updates in commit timestamp ($ct$) order.
A prepared transaction can commit with $ct$ bigger than the prepared time
proposed by the partition. 
A local stable time $stable_n$ is set to be smaller than the minimum
prepared time of the transactions prepared at the partition server
(Line~\ref{algpminprepared}).
To ensure updates are sent in $ct$ order, only updates with $ct \leq stable_n$
are sent to sibling partitions (Line \ref{algppropcausal}).
If {\small to\_send} is empty, 
the partition sends a heartbeat message (Line~\ref{heartbeat}). 
\item Each partition server maintains a vector \emph{$vec_p$} with an entry
  per site. An entry $j$ indicates that partition $p$ of site $i$
has delivered locally all updates with commit time
$ct \leq vec_p[j]$ from its sibling partition at site $j$. 
The local entry of the vector is set to $stable_n$,
as the $ct$ of prepared transactions is not defined.
  \item Periodically, each partition server sends its \emph{$vec_p$} to the other partition
  servers of its site (Line~\ref{algpsendst}).
 \item each partition $p_m^n$ computes $SV_m^n$, the latest stable snapshot known by $p_m^n$,
 as the minimum of all \emph{$vec_p$} received (Line~\ref{algpstablecompute}). 
  \end{enumerate}
  
  This protocol ensures that the snapshot is stable since 
  \begin{inparaenum}[(i)]
  \item no local partition will commit a transaction
  with commit time smaller than $vec_p[i]$, and 
  \item no remote update will be received 
  from a remote site $j$ with commit time smaller than $vec_p[j]$.
  \end{inparaenum}


\section{Evaluation}
\label{sec-evaluation}
We empirically explore how the results of the three-way trade-off
in Section~\ref{sec-tradeoff}
affect real workloads.
We evaluate \SYSTEM{} and the three minimal-delay protocols presented in
Section~\ref{sec-protocols}: CV, OP and AV.
%

\subsection{Implementation}

All protocols were built on the 
Antidote database \cite{antidote}, 
an open-source platform built 
using the Erlang programming 
language.
Antidote uses the Riak-core 
distributed hash table (DHT)
to partition the key-space 
evenly across physical servers \cite{riak}.
Object versions are stored in a per-partition in-memory
key-value store.
During the evaluation, updates were not persisted
to durable storage.

Under OP, AV and Cure, a linked list of recent updates is stored
for each key.
Old versions are garbage collected by a naive
mechanism that truncates version lists that contain 
more than 50 versions, keeping the latest 20.

\subsection{Setup}
\noindent\textbf{Hardware.} 
All experiments were run
on a cluster located in Rennes, France, on the 
Grid5000 \cite{grid5000} experimental platform
using fully-dedicated servers, where 
each server consists of 2 CPUs Intel Xeon E5-2630 v3,
with 8 cores/CPU, 128\,GB RAM, and two 558\,GB 
hard drives.
Nodes are connected through shared 10\,Gbps switches.
The ping latency measured within the cluster during the experiment
was approximately $0.15$ ms.

\smallskip\noindent\textbf{Configuration.}
Within the cluster, we configured two logical sites
consisting of 16 machines each.
Each site is comprised of 512 logical partitions, 
scattered evenly across the physical machines.
Nodes within the same DC communicate using 
the distributed message passing
framework of Erlang/OTP running over TCP.  
Connections across separate DCs
use ZeroMQ sockets \cite{zmq}, also over TCP,
where each node connects to
all other nodes to avoid any centralisation 
bottleneck.
The stabilisation protocol is run every 10\,ms
for OP, AV, and Cure.
This value has a negligible impact in throughput and
freshness, as noted as well in other systems
with similar stabilisation mechanisms
 \cite{rep:pro:sh182, db:syn:1752}.

\smallskip\noindent\textbf{Workload generation.}
The data set used in the experiments consists of 
100k keys per server, totalling
1.6 million keys.
Objects are registers with the last-writer wins (LWW) policy \cite{db:rep:optim:1454},
where updates generate random 100-byte binary values.
All objects were replicated at all sites.
A custom version of Basho Bench
is used to generate the workload \cite{basho_bench}.
Google's Protocol Buffer interface is used to serialise messages between
Basho Bench clients and Antidote servers \cite{buffers2011google}.
To avoid across-machine latencies, two instances
of Basho Bench run at each server, which issue requests to the Antidote
instance running on it.
Each run of the benchmark was run for two minutes, the first minute
being used as a warm-up period.
A variable number of clients repeatedly run read-only and update
transactions.
We run first instances with high update rate and number of clients.  
This rapidly populates the store up-to a point where memory utilisation 
remains constant due to garbage collection, and response times stabilise.
Operations within a transaction
select their keys using a power-law distribution, where 80\%
of the operations are directed to 20\% of the keys.

\subsection{Experiments}

We expect to observe a similar latency response
for all minimal-delay protocols, and a slight degradation for Cure,
which may block for a small amount of time under clock skew or due to
reading concurrent updates.
We expect to observe lower latency for CV, as it does not incur the 
overheads of multi-versioning.
Moreover, we expect to observe
Order-Preserving visibility to exhibit significantly better freshness than 
Cure and AV, which implement Atomic Visibility.
We run each experiment in two configurations: 
the cluster configured as a single DC, or logical split into 
two DCs.
We observed very similar results under both configurations.
In what follows, we only present the results of the
latter.

\smallskip\noindent\textbf{Workloads.}
We run experiments under two workloads which run different 
read-only transactions. 
Under the first workload, a read-only transaction reads a
number of objects in parallel,
in a single round, i.e., in a single call.
In the second workload, we try to mimic Facebook's multi-read operations,
by issuing read-only transactions that make many calls, each reading
a number of objects in parallel.
Under both workloads, each client repeatedly executes a read-only transaction 
followed by an update transaction in a closed-loop (zero think time).
An update transaction performs blind updates over a subset of the objects read by
the previous transaction. 
Under every evaluated protocol, this is equivalent to running a single transaction
that executes reads followed by updates. 
We use the separation to simplify the latency measurement of a transaction's read phase.

We vary the number of client threads and measure how latency,
and freshness change as load is added to the system.
Throughput is measured in operations per second, where each operation is a read
or write in a transaction. 
We measure staleness as follows.
When a partition server receives a read request, it logs asynchronously
the number of versions needed to skip to guarantee the required isolation property.
For Cure, it also logs the cases where it has to wait due to clock skew or 
for other transactions to commit. 
We process these logs offline.

\smallskip\noindent\textit{Single-shot read-only transactions.}
A single-round transaction performs 100 reads in
parallel. Update transactions perform 
two, 10 or 100 updates, generating an update rate of approximately
2, 9 and 50\% respectively.

\smallskip\noindent\textit{Multi-shot read-only transactions.}
The multi-round experiment mimics the Facebook social network, where a
transaction reads thousands of objects in tens to dozens of rounds, and
updates represent 0.2\% of the workload \cite{tao}.
In this workload, a read-only transaction executes
10 rounds of 100 parallel reads each (totalling 1000 reads).
An update transaction performs two, 100 or 1000 updates, 
generating an update rate of approximately 0.2, 9 and 50\% respectively.

\subsubsection{Throughput results.}
Due to space limitations, we do not include throughput figures.
We explain briefly the behaviour we observed during both experiments.
Graphs and a detailed explanation can be found in an extended version of this work \cite{thesis-ale}. 

CV exhibits the highest throughput. 
The other protocols must manage multiple versions; 
since CV only requires the most recent version, it does not have this overhead.
All other protocols exhibit throughput similar to one another throughout the evaluation space.
The difference in throughput, between CV and the other protocols, is negligible for workloads with a low rate of updates.  
It grows to 15\% for a workload with $\approx$10\% of updates, and to 35\% for 50\% of updates.
Under Cure, AV and OP, a higher proportion of updates increases
(i) the frequency of garbage collection, and 
(ii) the number of updates running concurrently with a read operation.
The latter increases the overhead of traversing version lists to find
a version that satisfies a given snapshot.

\subsubsection{Single-shot read-only transactions}
In this section, we present the results of latency and freshness of the first experiment. 

\begin{figure}[t!]
\centering
	\includegraphics[width=.6\linewidth]{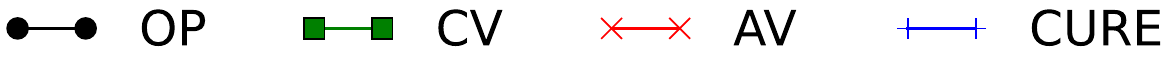}
    
	\begin{subfigure}{0.32\columnwidth}
	  \centering
	  \includegraphics[width=\linewidth]{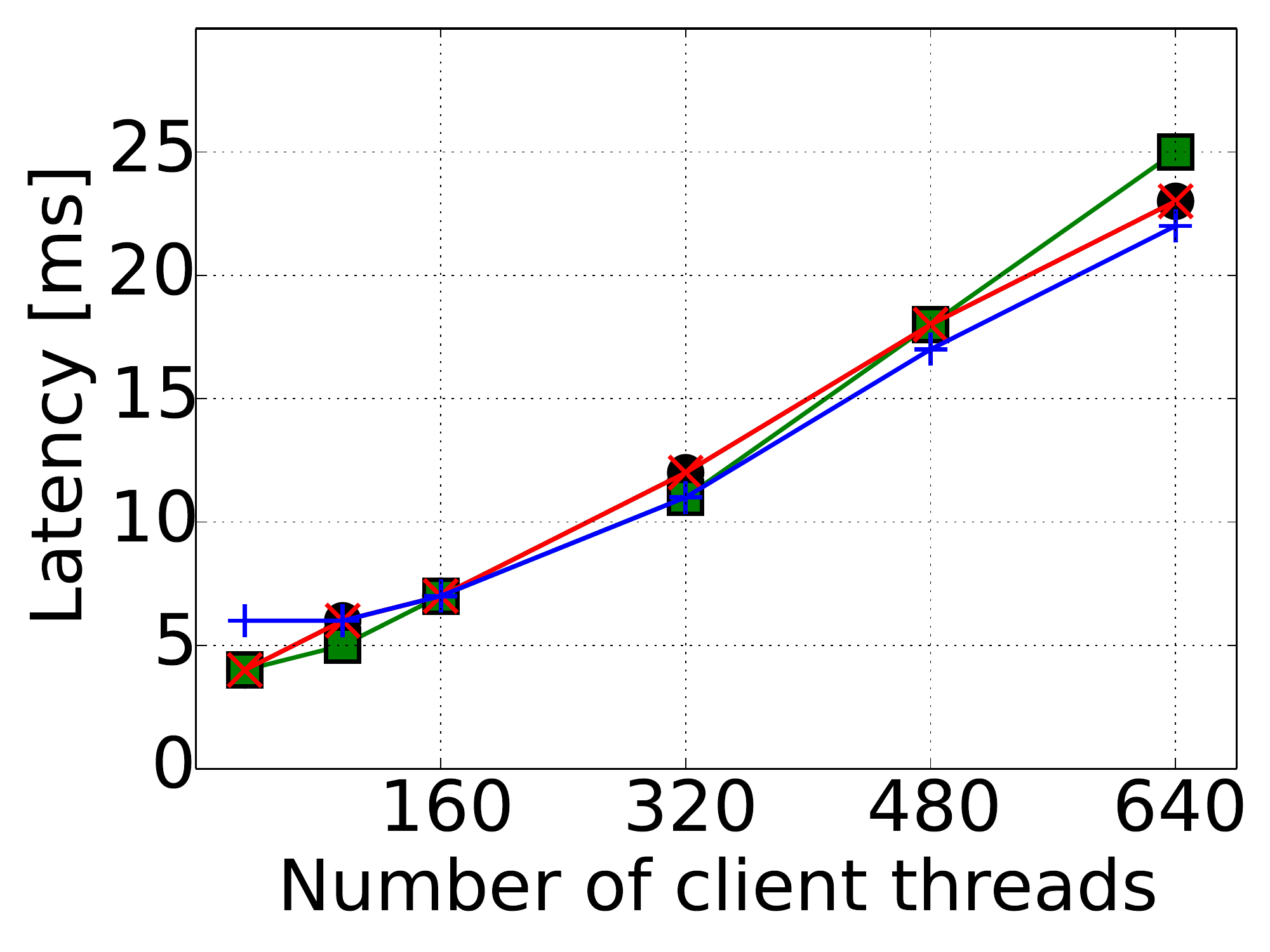}
	  \caption{latency w:2}
	  \label{fig:one-lat-2}
	\end{subfigure}%
	\begin{subfigure}{0.32\columnwidth}
	  \centering
	  \includegraphics[width=\linewidth]{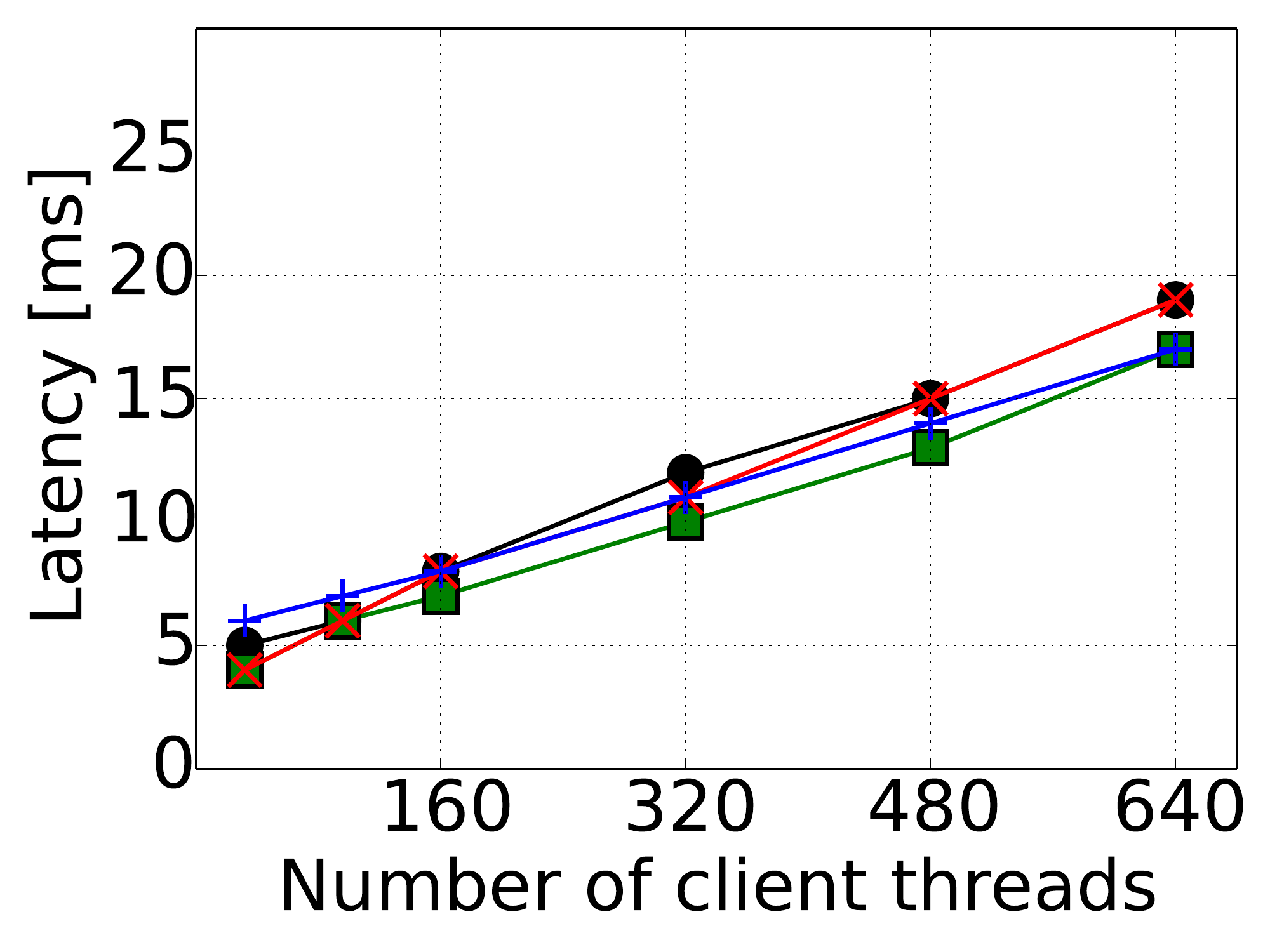}
	  \caption{latency w:10}
	  \label{fig:one-lat-10}
	\end{subfigure}%
	\begin{subfigure}{0.32\columnwidth}
	  \centering
	  \includegraphics[width=\linewidth]{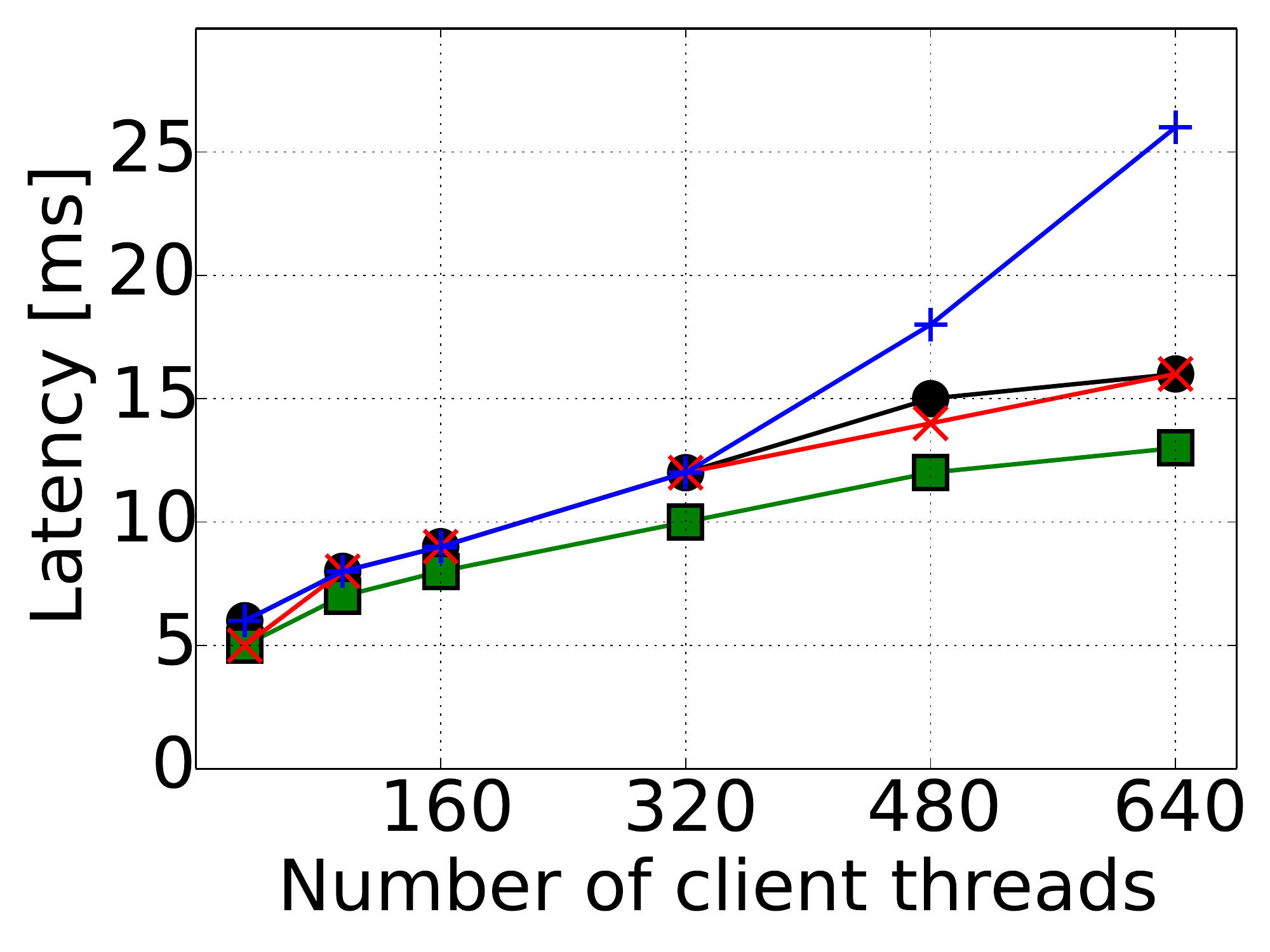}
	  \caption{latency w:100}
	  \label{fig:one-lat-100}
	\end{subfigure}%

\begin{subfigure}{0.33\columnwidth}
  \centering
  \includegraphics[width=\linewidth]{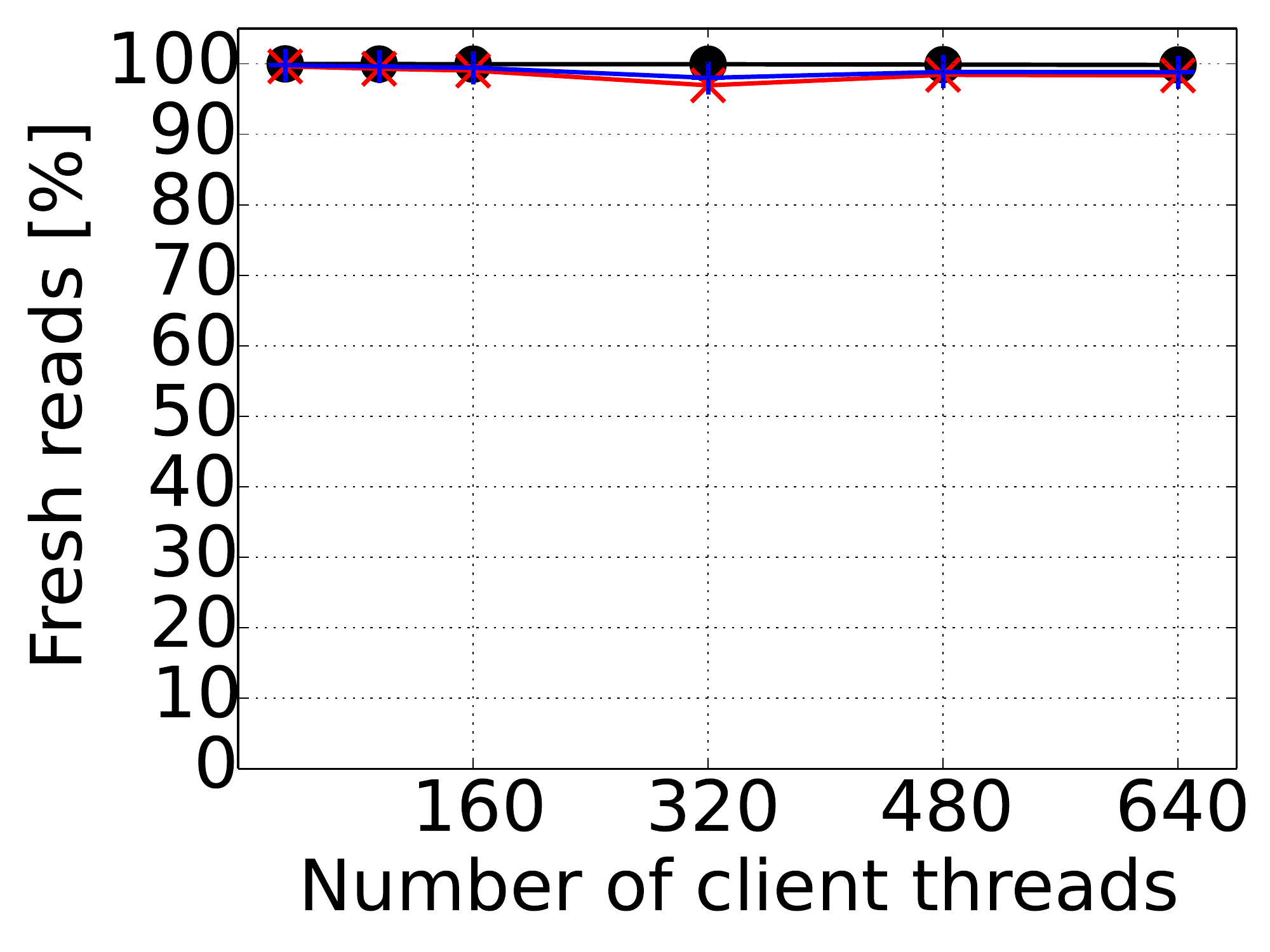}
  \caption{freshness w:2}
  \label{fig:one-fresh-2}
\end{subfigure}
\begin{subfigure}{0.32\columnwidth}
  \centering
  \includegraphics[width=\linewidth]{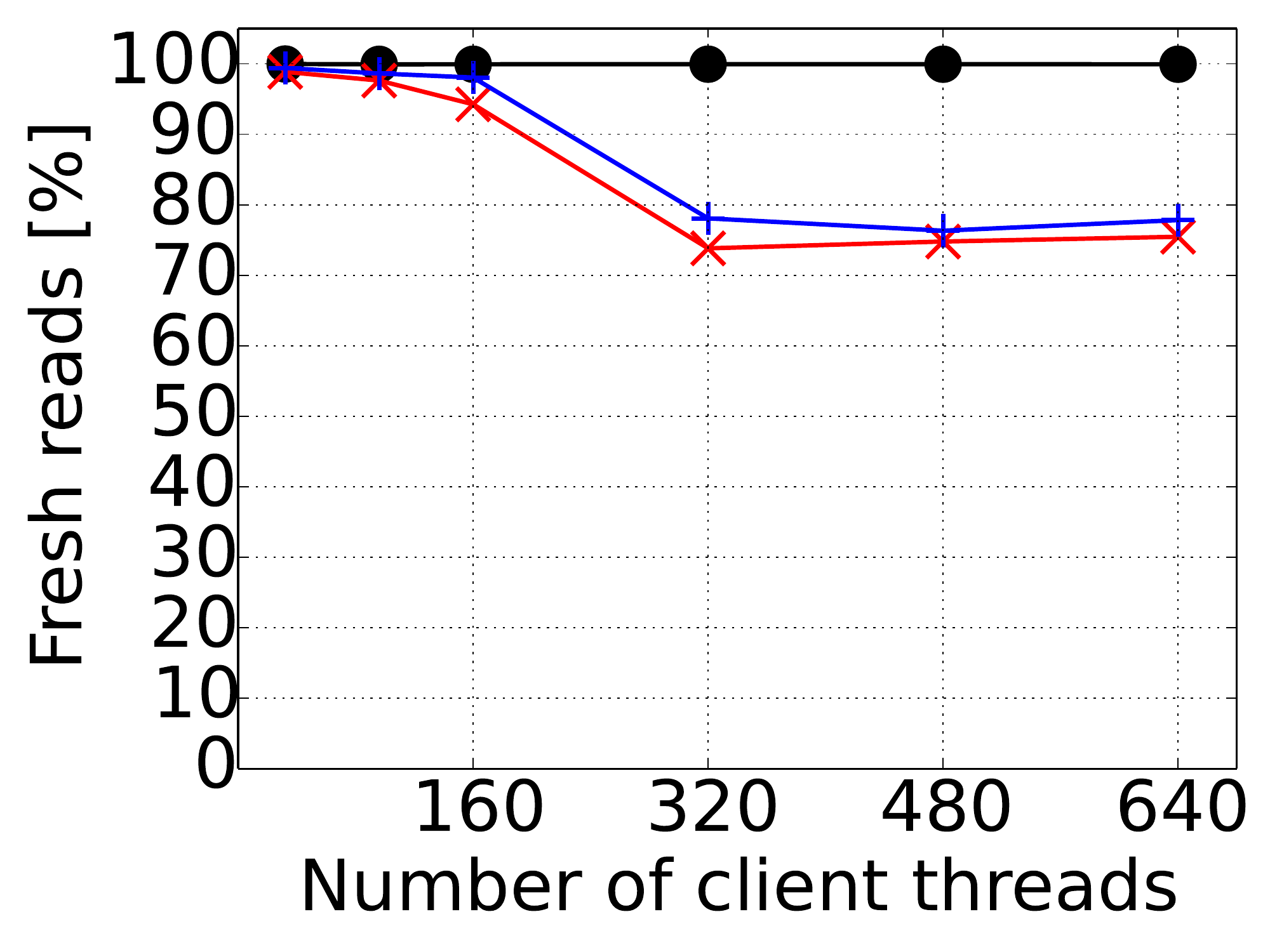}
  \caption{freshness w:10}
  \label{fig:one-fresh-10}
\end{subfigure}
\begin{subfigure}{0.32\columnwidth}
  \centering
  \includegraphics[width=\linewidth]{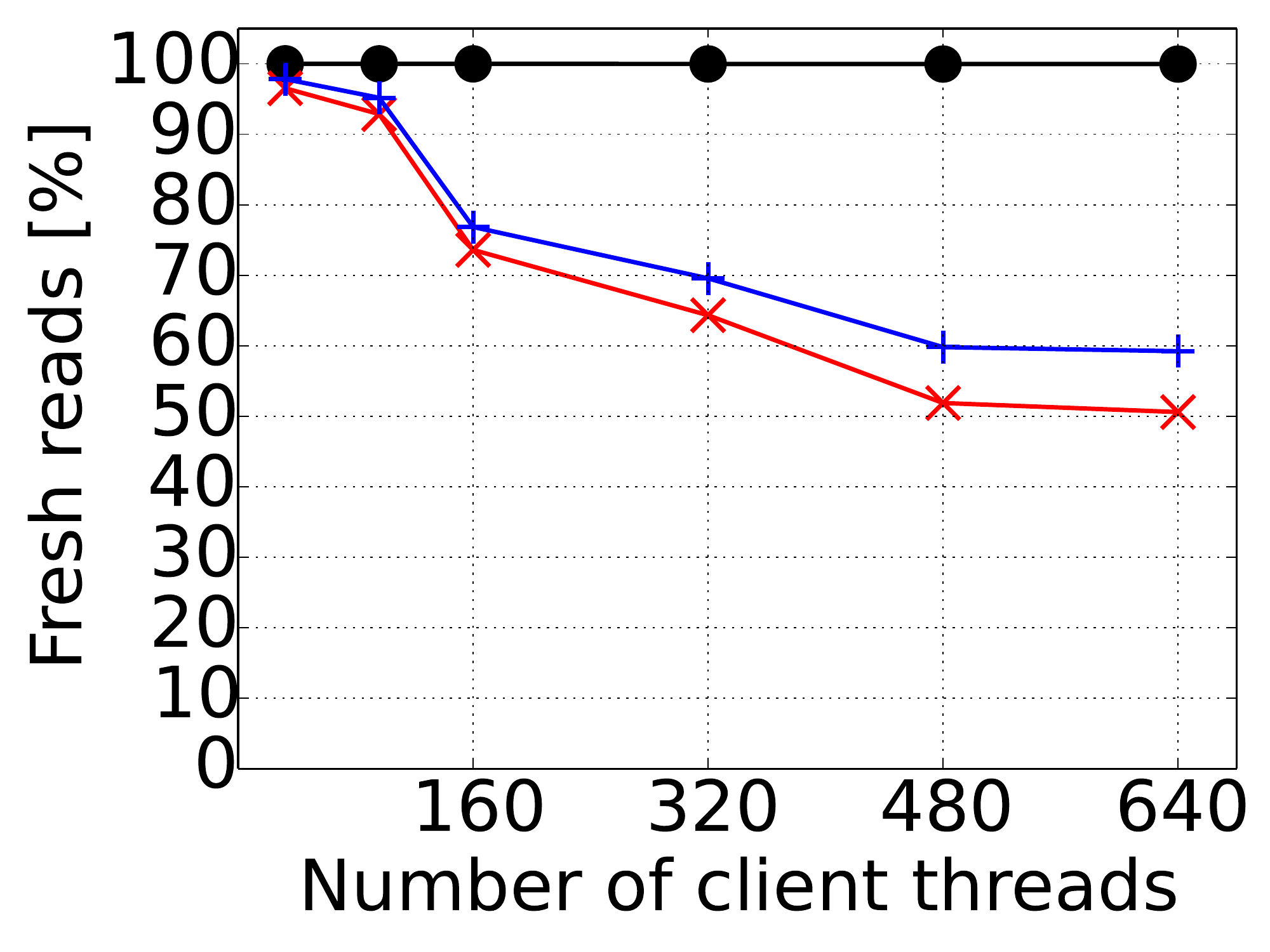}
  \caption{freshness w:100}
  \label{fig:one-fresh-100}
\end{subfigure}

\begin{subfigure}{0.32\columnwidth}
  \centering
  \includegraphics[width=\linewidth]{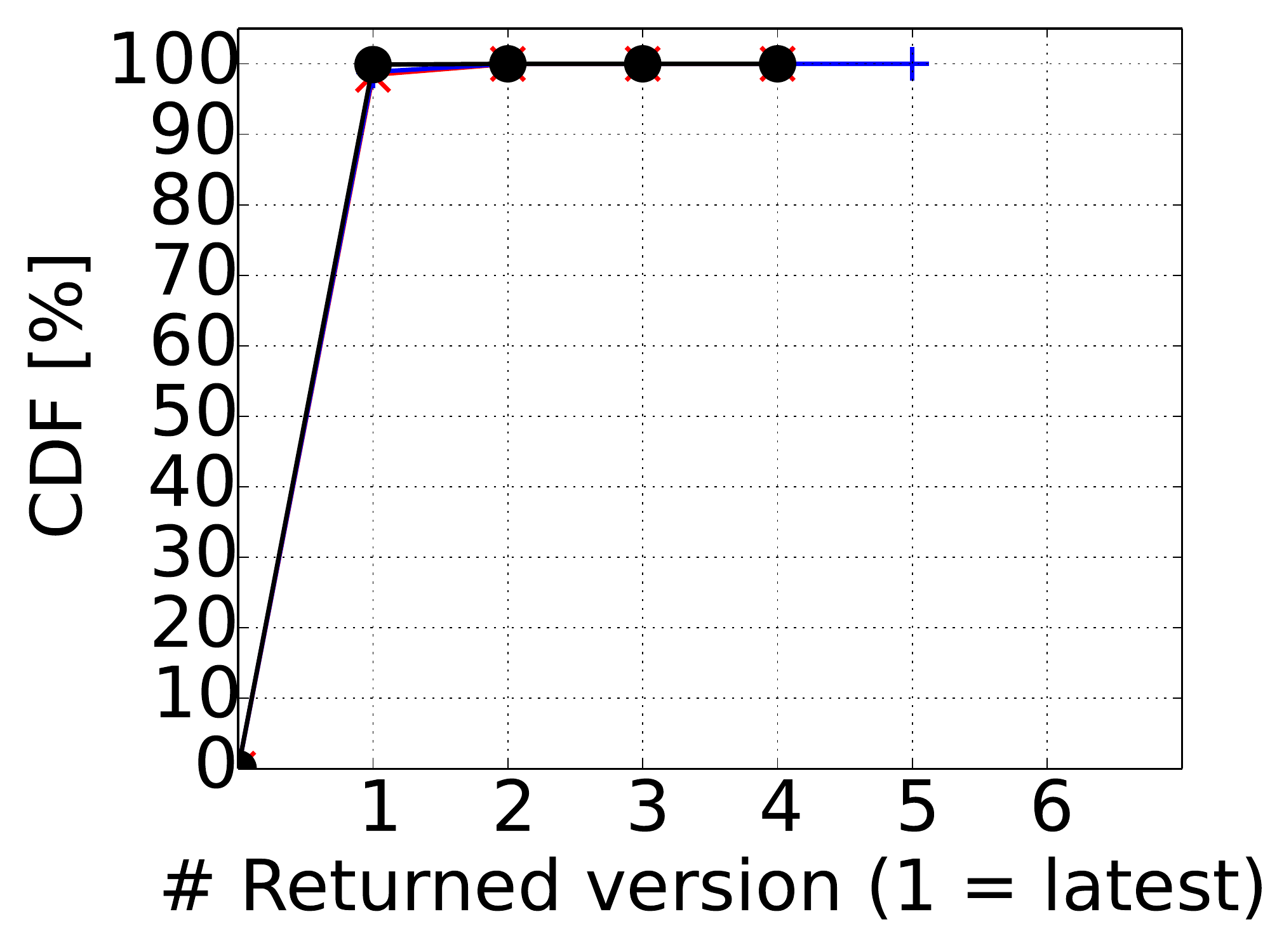}
  \caption{CDF w:2-th:480}
  \label{fig:one-cdf-2}
\end{subfigure}
\begin{subfigure}{0.32\columnwidth}
  \centering
  \includegraphics[width=\linewidth]{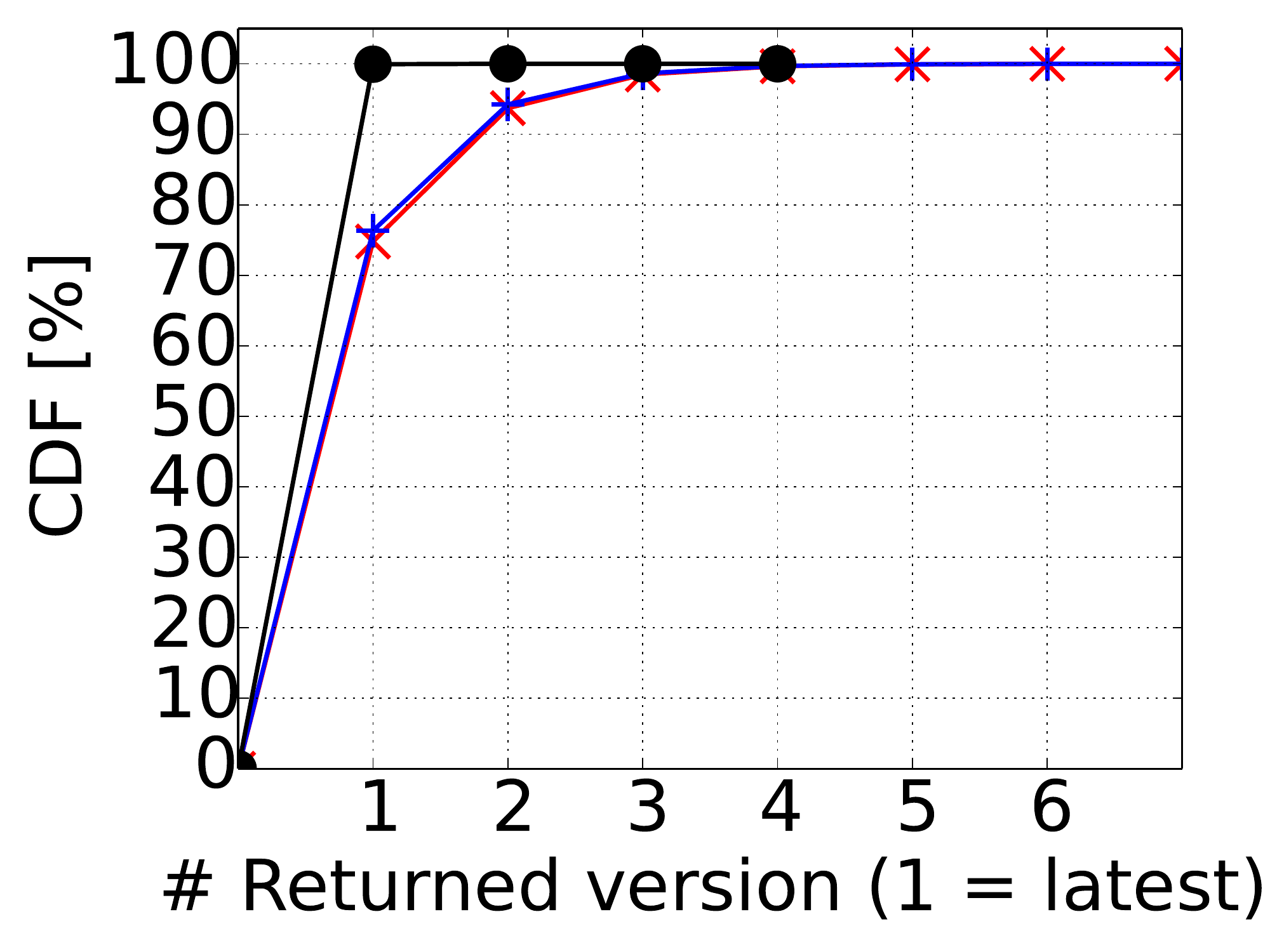}
  \caption{CDF w:10-th:480}
  \label{fig:one-cdf-10}
\end{subfigure}
\begin{subfigure}{0.32\columnwidth}
  \centering
  \includegraphics[width=\linewidth]{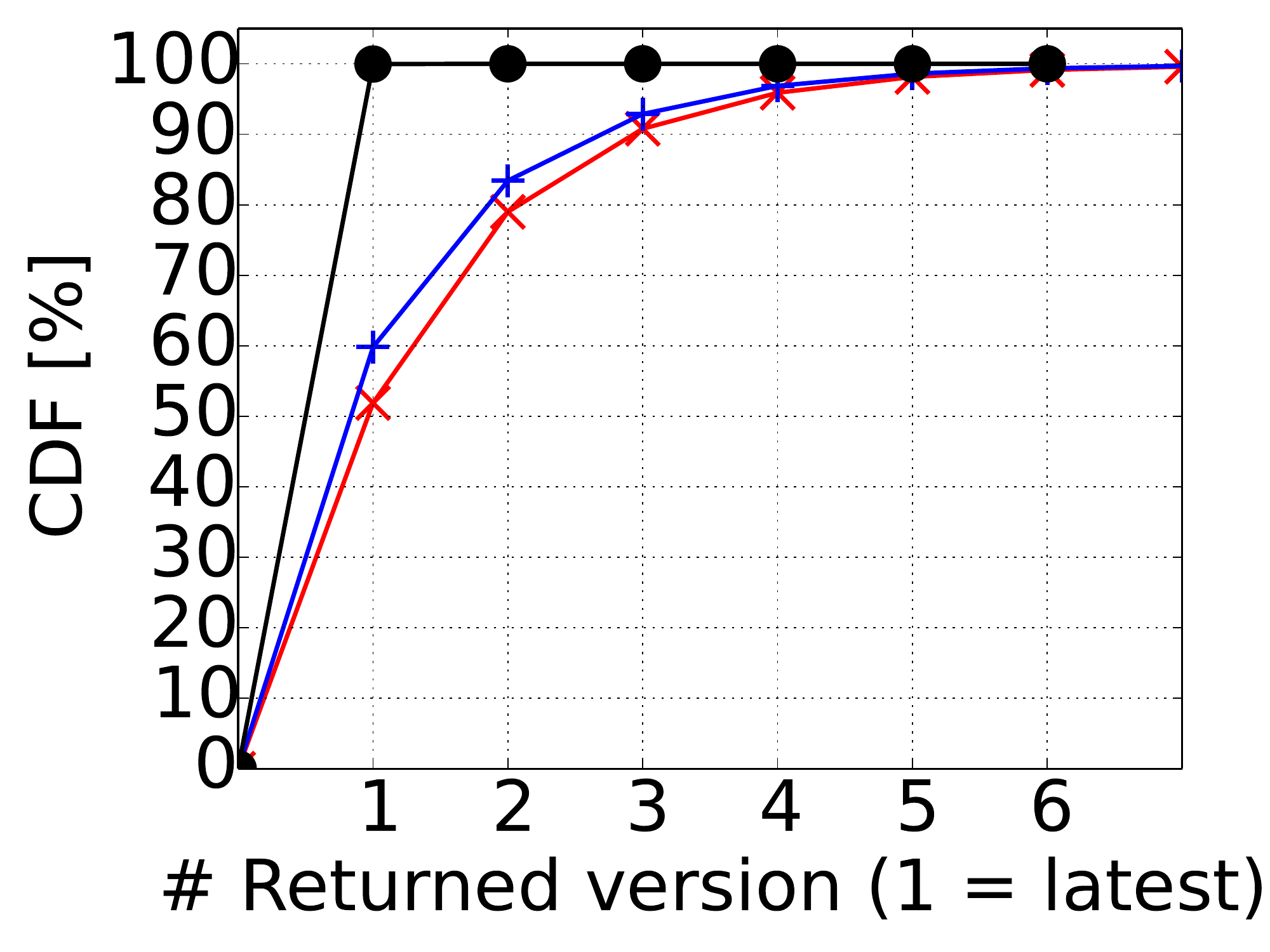}
  \caption{CDF w:100-th:480}
  \label{fig:one-cdf-100}
\end{subfigure}

\begin{subfigure}{0.32\columnwidth}
  \centering
  \includegraphics[width=\linewidth]{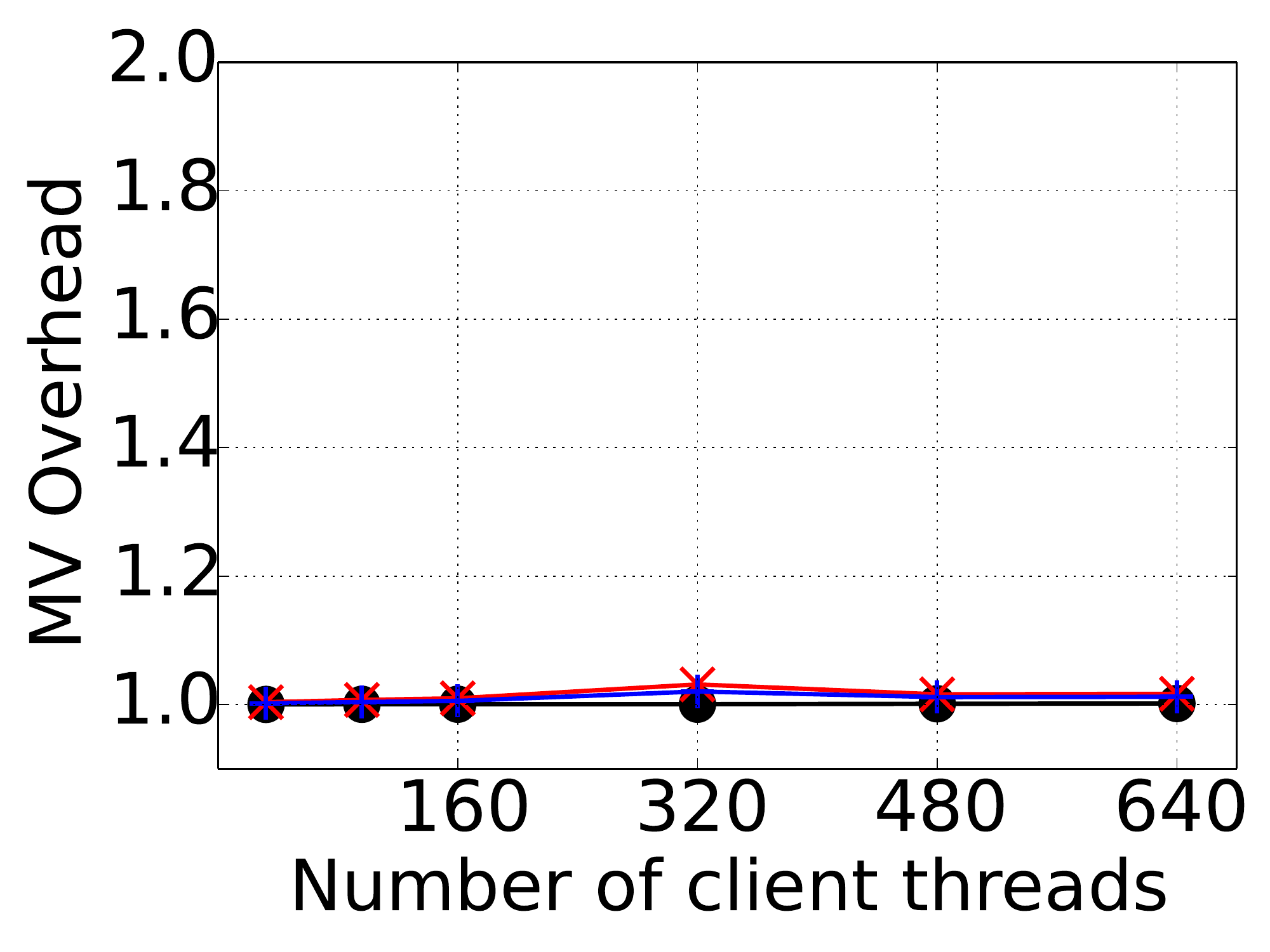}
  \caption{MV oh. w:2}
  \label{fig:one-moh-2}
\end{subfigure}
\begin{subfigure}{0.31\columnwidth}
  \centering
  \includegraphics[width=\linewidth]{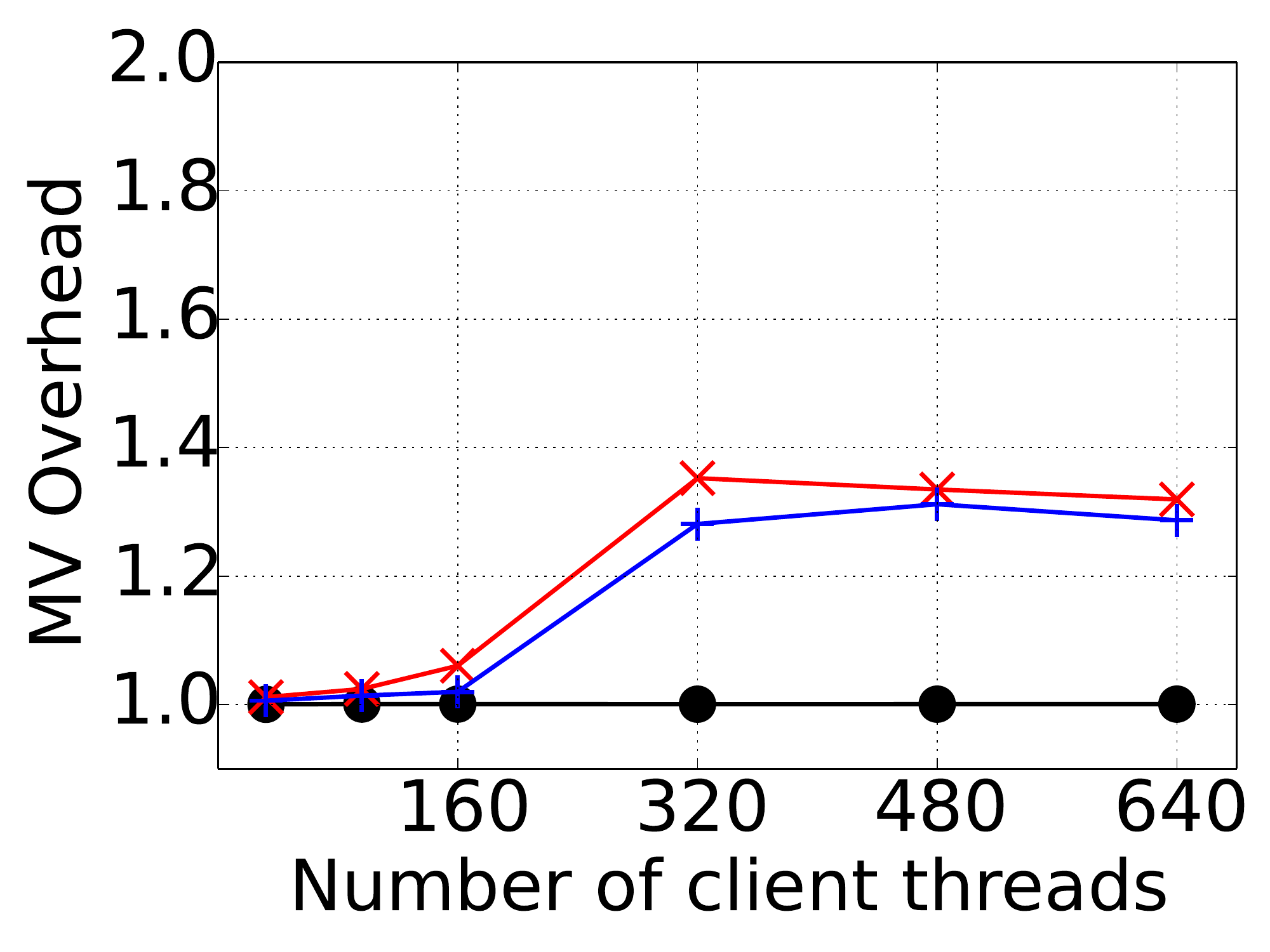}
  \caption{MV oh. w:10}
  \label{fig:one-moh-10}
\end{subfigure}
\begin{subfigure}{0.31\columnwidth}
  \centering
  \includegraphics[width=\linewidth]{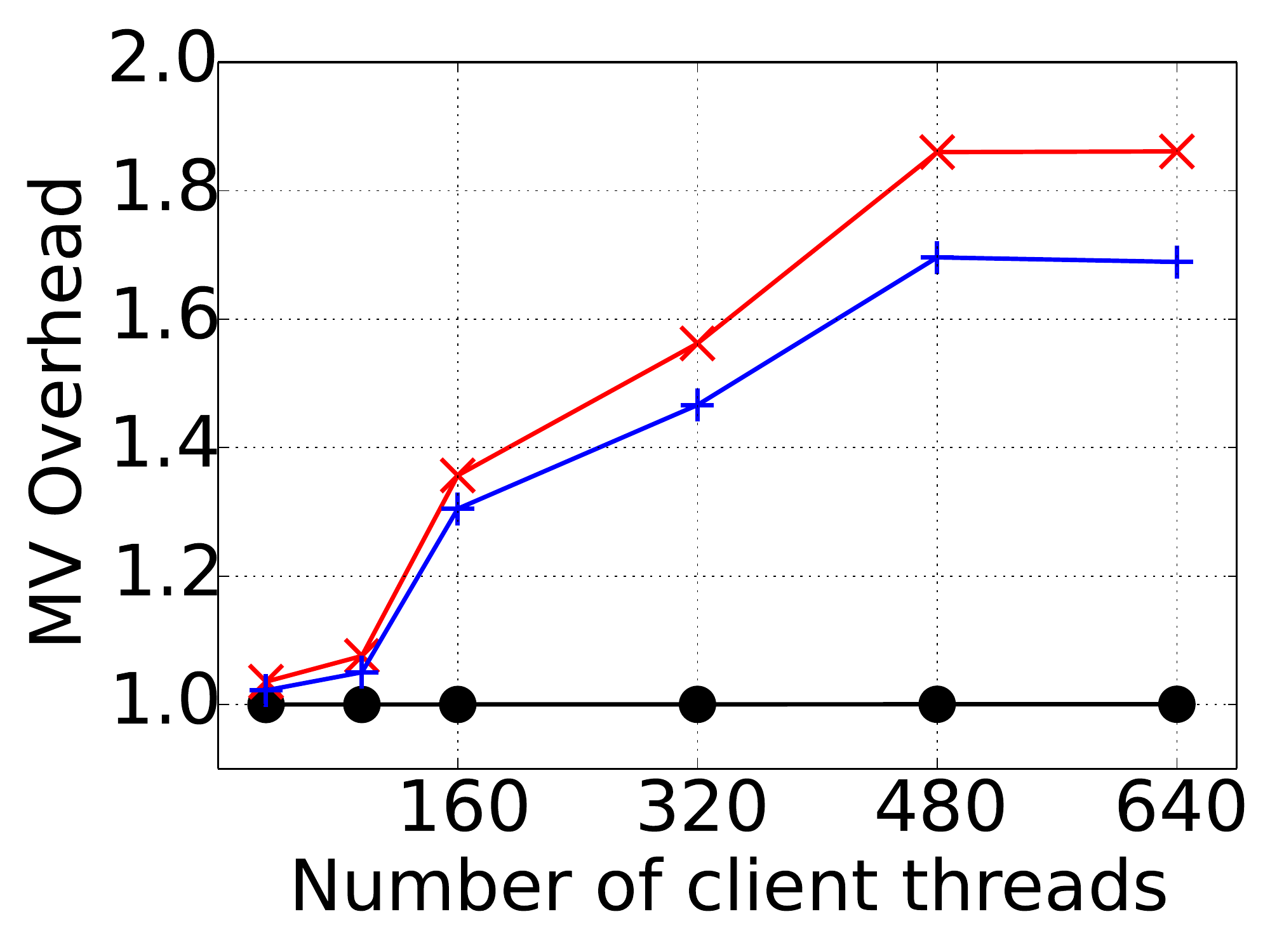}
  \caption{MV oh. w:100}
  \label{fig:one-moh-100}
\end{subfigure}

\caption{Single-shot read-only-transactions}
\label{fig:eval-single}
\end{figure}

\smallskip
\noindent\textbf{Latency.} 
Figures \textbf{\ref{fig:one-lat-2}}, \textbf{\ref{fig:one-lat-10}} and \textbf{\ref{fig:one-lat-100}}
show the latency of single-shot read-only transactions.
All protocols exhibit a similar trend: increasing the offered load increases
latency. 
CV, AV and OP exhibit minimal delays and, therefore, exhibit very similar latency. 
CV does not incur the overhead for searching for a compatible version, and 
has slightly lower latency than OP and AV.
Cure exhibits delays. For low update rates,
Figures \textbf{\ref{fig:one-lat-2}} and \textbf{\ref{fig:one-lat-10}} show that, under small
number of client threads, Cure exhibits extra (up to 1.9X) latency due to clock skew between servers.
At high update rate (Figure \textbf{\ref{fig:one-lat-100}}),
read operations in Cure wait for update transactions to commit frequently.
This causes the latency gap between this protocol and the remaining ones when the number
of client threads is large.
In Appendix \ref{app:cure-blocking}, we analyse these effects in more depth.

\smallskip\noindent\textbf{Freshness.}
Figures \textbf{\ref{fig:one-fresh-2}}, \textbf{\ref{fig:one-fresh-10}} and \textbf{\ref{fig:one-fresh-100}} show the freshness response 
as the number of client threads increases, for different update rates.
Plots display
the percentage of read operations that returned the most up-to-date
version available at the contacted server.
CV is not present in the figures as it always returns the latest
version.
Figures \textbf{\ref{fig:one-cdf-2}}, \textbf{\ref{fig:one-cdf-10}} and \textbf{\ref{fig:one-cdf-100}} display a CDF showing
how stale a read version is under 480 client threads for each update rate. 
We observe that:
\begin{itemize}[leftmargin=*]
\item OP outperforms the other protocols under all workloads.
Its freshness response remains nearly constant:
over 99.8\% of reads observe the latest version under all configurations.
This shows that Concurrent Freshness allows this protocol
to behave nearly optimally.
\item Under the 2\%-writes workload (Figure \textbf{\ref{fig:one-fresh-2}}), OP exhibits, 
0.2\% of stale reads in the worst case, while Cure 2\% (10X), and AV 3\% (15X).
Figure \textbf{\ref{fig:one-cdf-2}} shows how fresh reads were under 480 client threads.
All protocols read, most of the times, the latest or second most recent
version. OP read, in the worst case,  the third most recent
version, while the remaining two protocols, the fourth.
Cure exhibits, under the same conditions,  1.2\% (6X) and, AV, 1.8\% (9X) stale reads, 
meaning that potentially every transaction observes stale versions.
\item Under 10\% of writes (Figure \textbf{\ref{fig:one-fresh-10}}),
OP does not further degrade
its freshness, and reads observe the same percentage of stale versions,
whereas freshness degrades significantly for AV,
which shows 25\% (125X) stale reads in the worst case, and Cure, which shows  22\% (110X).
Figure \textbf{\ref{fig:one-cdf-10}} shows that OP read mostly fresh versions and,
in the worst case,  the fourth most-recent
version. Cure and AV show a higher frequency or returning the second ($\approx$ 20\%),
third ($\approx$ 5\%)
			and fourth ($\approx$ 2\%) most recent versions. As each read transaction
			executes 100 reads, this means that this
potentially affects every transaction.
In the worst case, these two protocols read the 12th-to-latest version,
not shown in the picture to display them more clearly.
\item Under 50\% of writes (Figure \textbf{\ref{fig:one-fresh-100}}),
we observe  the biggest difference
between all protocols: 
OP still suffers from up to 0.2\% of stale reads, while Cure from up to
41\% (205X) and AV 49\% (245X).
Figure \textbf{\ref{fig:one-cdf-100}} shows that, for 480 threads OP read,
in the worst case, the 6th-to-latest version.
Cure the 18th and AV the 19th.
Cure and AV frequently show significantly stale versions, up to the sixth
($\approx$ 2\%) version is potentially observed
by every transaction.  The oldest version returned by AV was the 19th to latest,
and by Cure, the 18th to latest.
\end{itemize}

%


\smallskip\noindent\textbf{Multi-version overhead.}
We compute the multi-version overhead as the extra work required, for a read operation,
to find and store a version that respects a required isolation level, 
with respect to a single-versioned protocol (e.g., CV).
For instance, under this metric, a read that returns the second-to-latest version has an overhead of 2X
over the baseline.
This metric is computed in practice as the area over the lines of the CDFs. 
We compute this metric as the overall overhead observed during the entire execution. 
Figures \textbf{\ref{fig:one-moh-2}}, \textbf{\ref{fig:one-moh-10}} and \textbf{\ref{fig:one-moh-100}} show the results under this workload.

For all workloads, OP shows a very low overhead, under 1.002X over an optimal
protocol.
AV presents a maximum overhead of 1.03X, 1.35X, and 1.87X under 2, 10 and 100
updates per transaction, respectively,
while Cure 1.02X, 1.31X and 1.7X.

\smallskip\noindent\textbf{Conclusion.}
We have observed the effects of the three-way trade-off
in action.
Under this workload, strengthening the semantics from Committed
to Order-Preserving visibility incurs a negligible overhead in terms of latency and freshness.
However, strengthening the semantics to Atomic Visibility penalises freshness significantly. 
Both protocols we have experimented with exhibited a high degradation in freshness.
Cure exhibits better freshness than AV at a latency cost that increases with contention.

\subsubsection{Multi-shot read-only transactions.}

\begin{figure}[t!]
\centering
\hspace*{3mm}
	\includegraphics[width=.6\linewidth]{figures/legend.pdf}
	
	\begin{subfigure}{0.32\columnwidth}
	  \centering
	  \includegraphics[width=\linewidth]{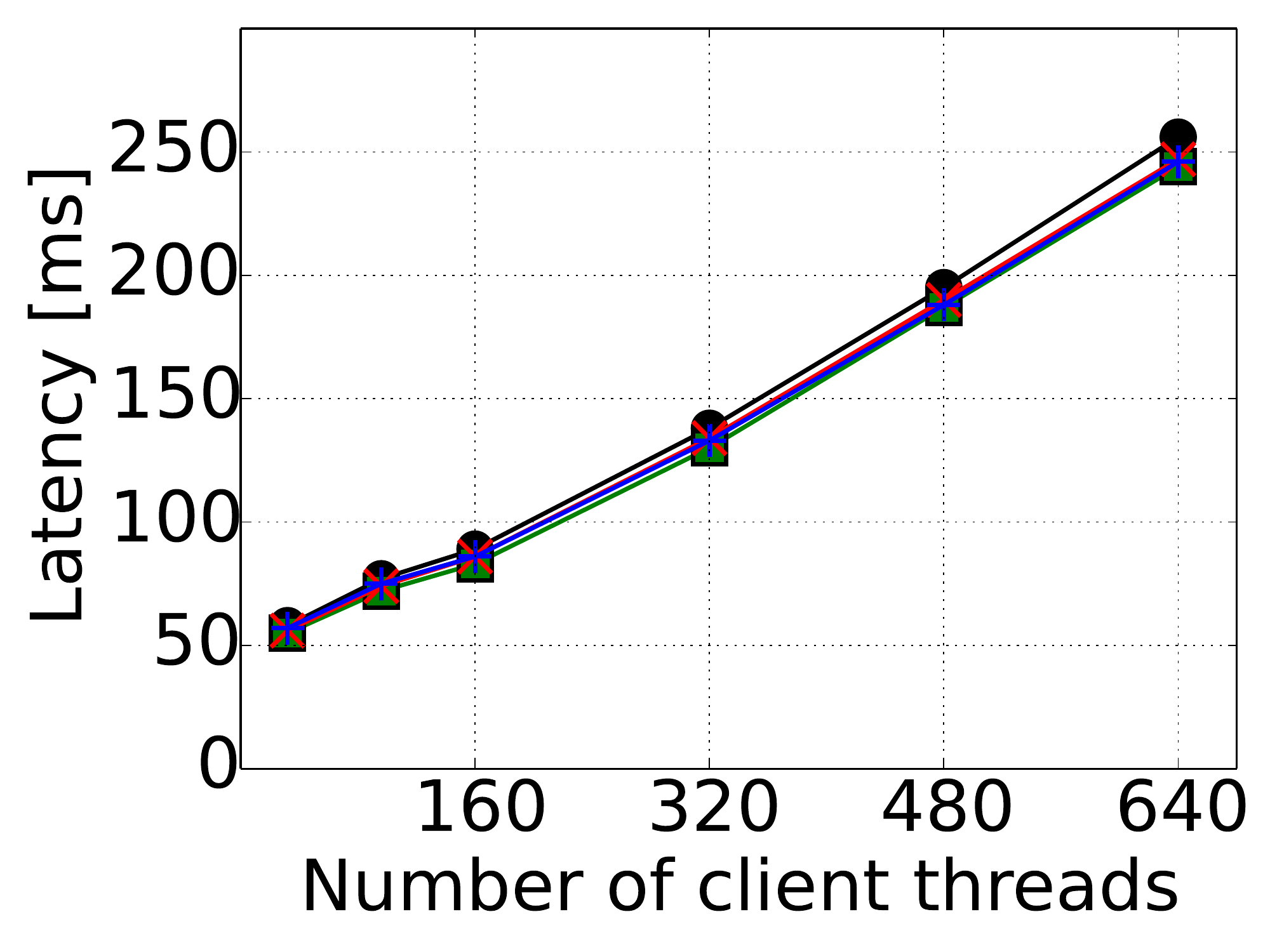}
	  \caption{latency w:2}
	  \label{fig:face-lat-2}
	\end{subfigure}%
	\begin{subfigure}{0.32\columnwidth}
	  \centering
	  \includegraphics[width=\linewidth]{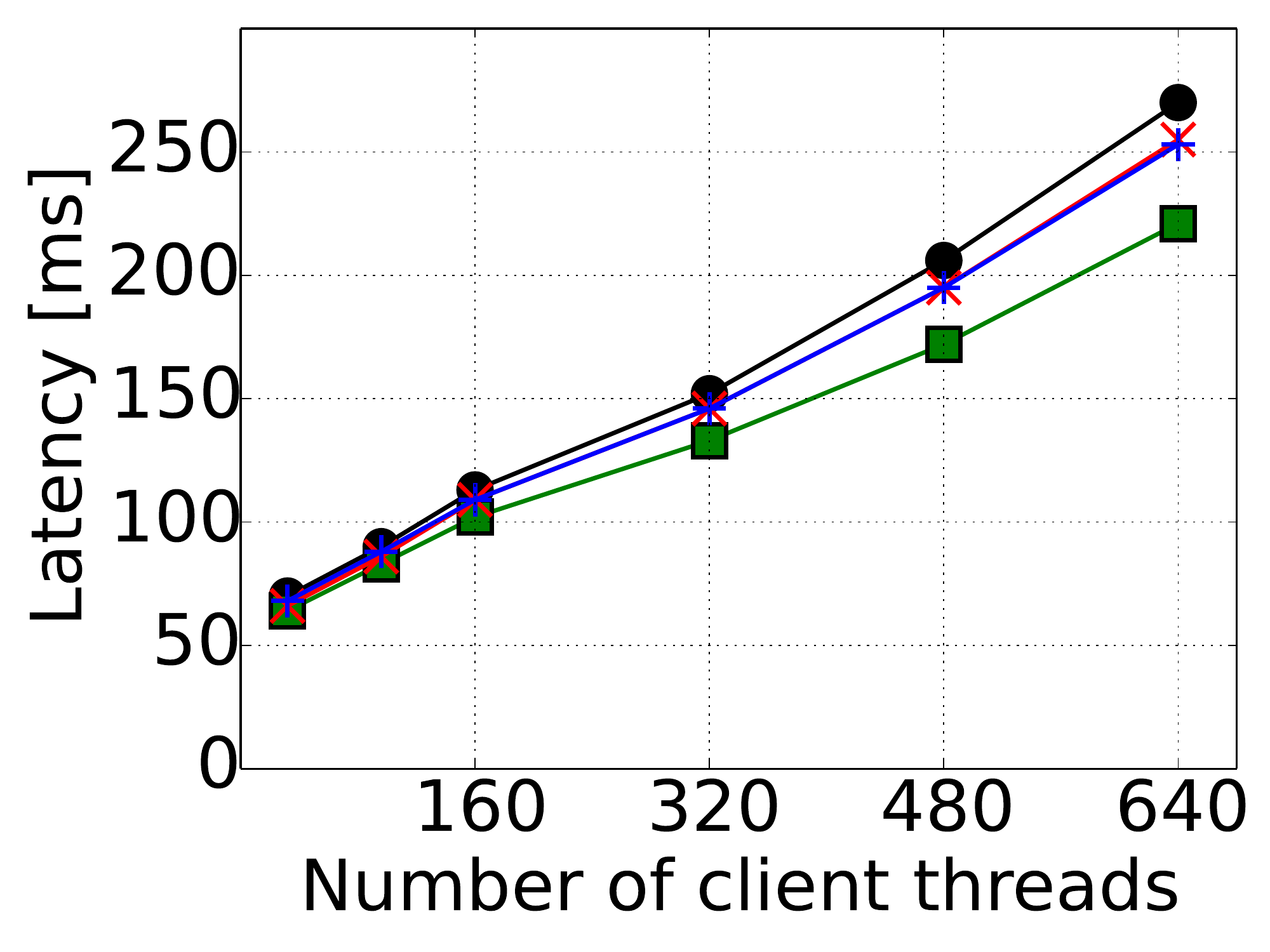}
	  \caption{latency w:100}
	  \label{fig:face-lat-100}
	\end{subfigure}%
	\begin{subfigure}{0.32\columnwidth}
	  \centering
	  \includegraphics[width=\linewidth]{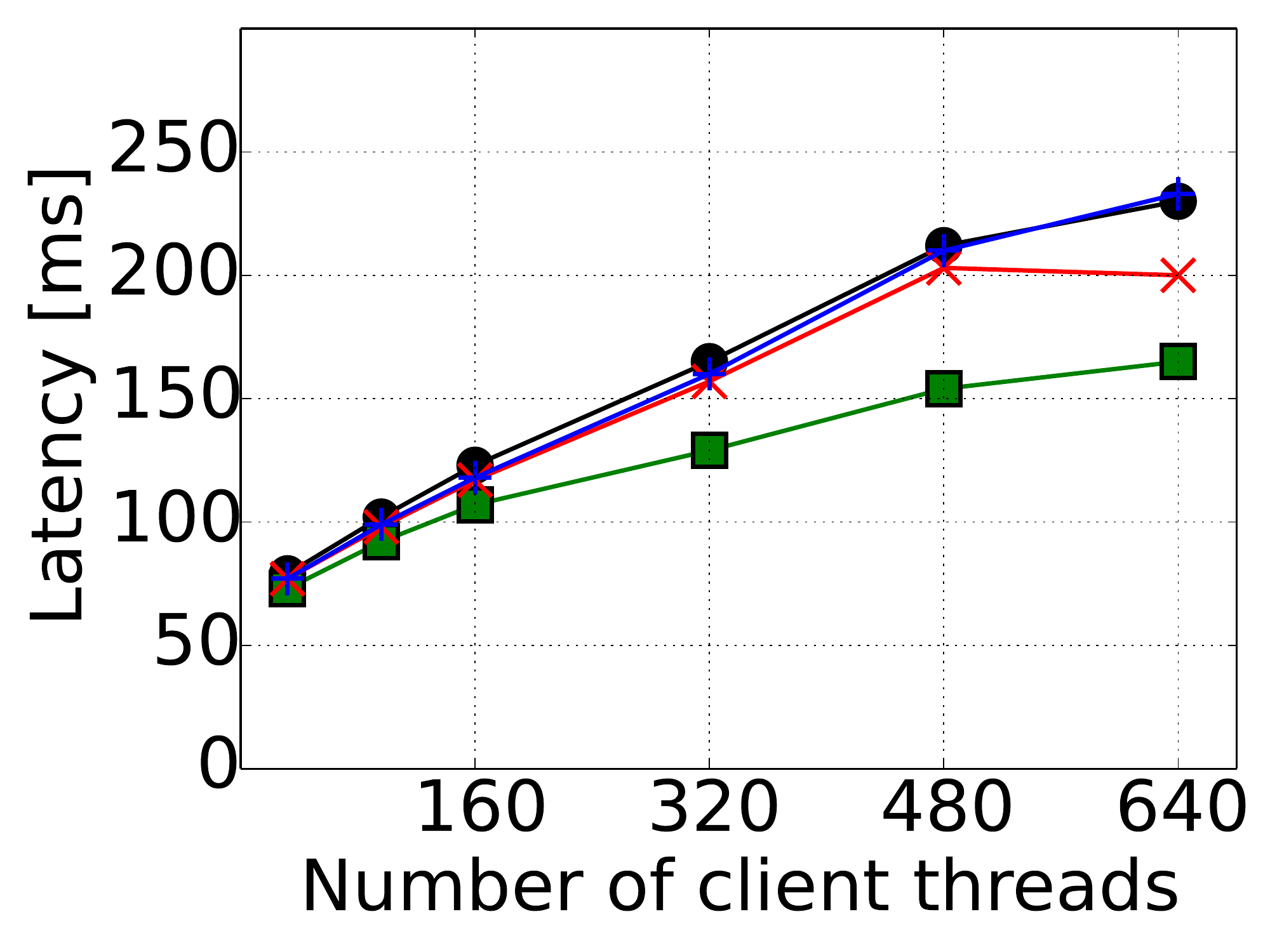}
	  \caption{latency w:1000}
	  \label{fig:face-lat-1000}
	\end{subfigure}%

\begin{subfigure}{0.33\columnwidth}
  \centering
  \includegraphics[width=\linewidth]{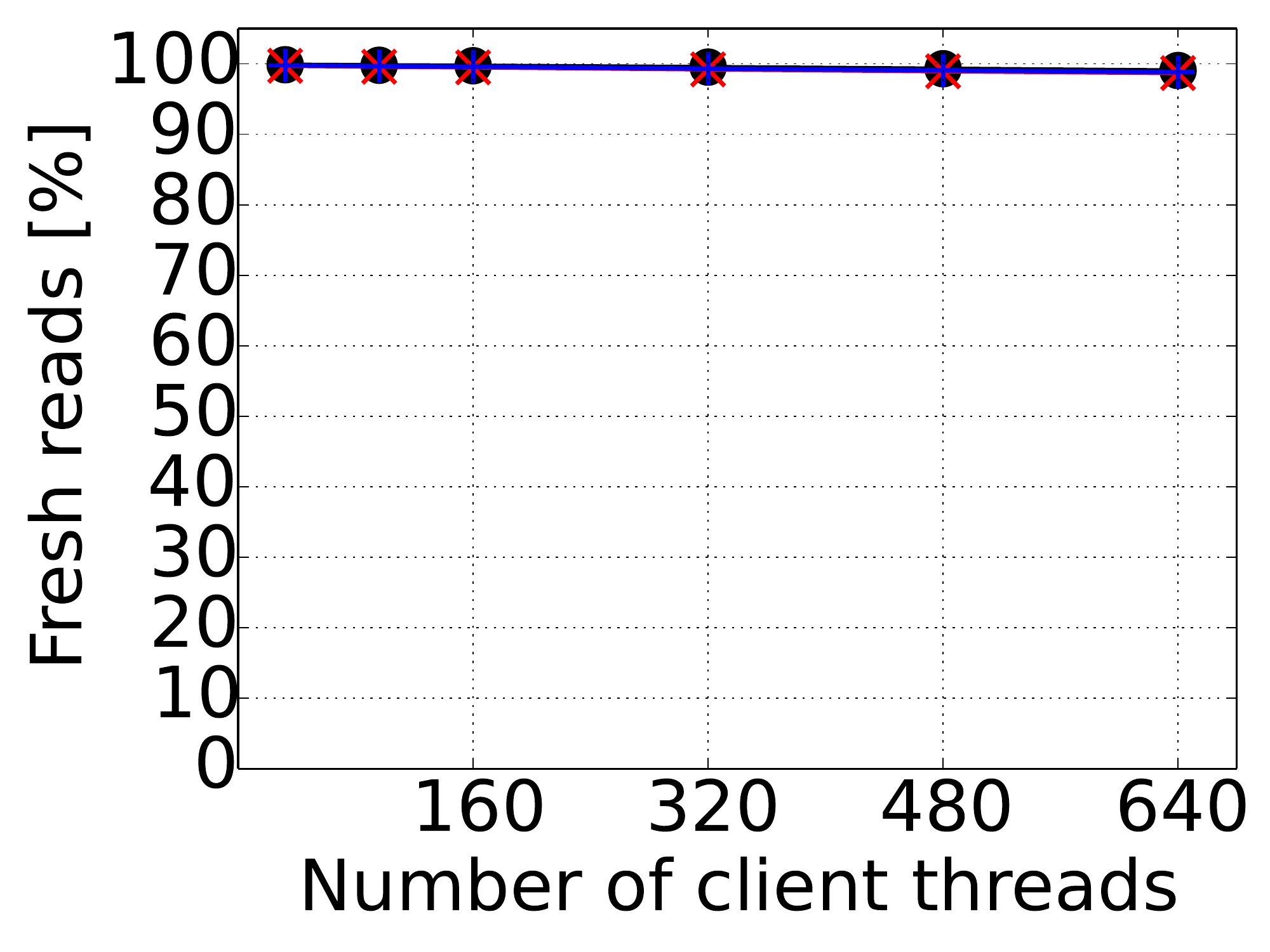}
  \caption{freshness w:2}
  \label{fig:face-fresh-2}
\end{subfigure}
\begin{subfigure}{0.32\columnwidth}
  \centering
  \includegraphics[width=\linewidth]{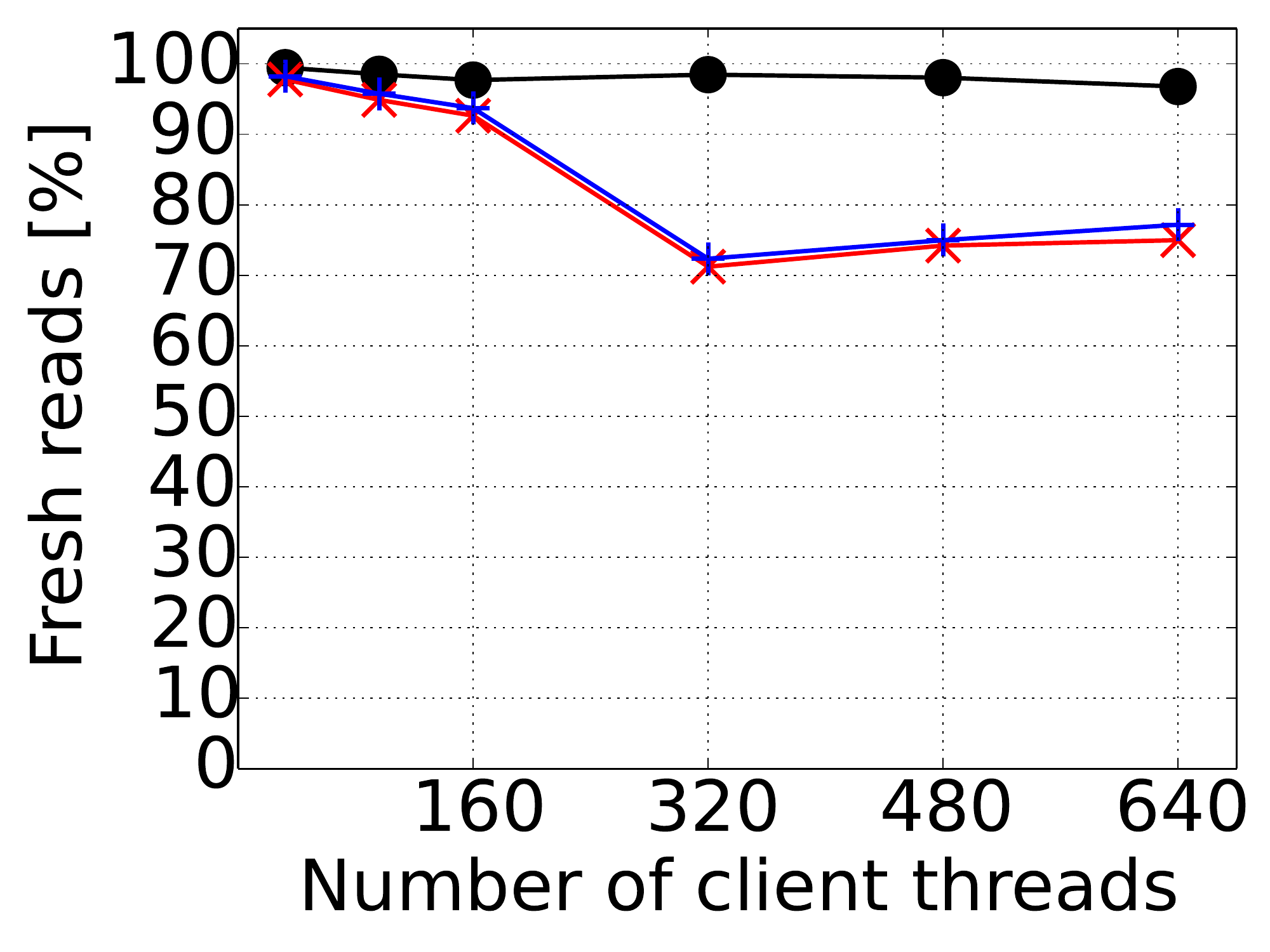}
  \caption{freshness w:100}
  \label{fig:face-fresh-100}
\end{subfigure}
\begin{subfigure}{0.32\columnwidth}
  \centering
  \includegraphics[width=\linewidth]{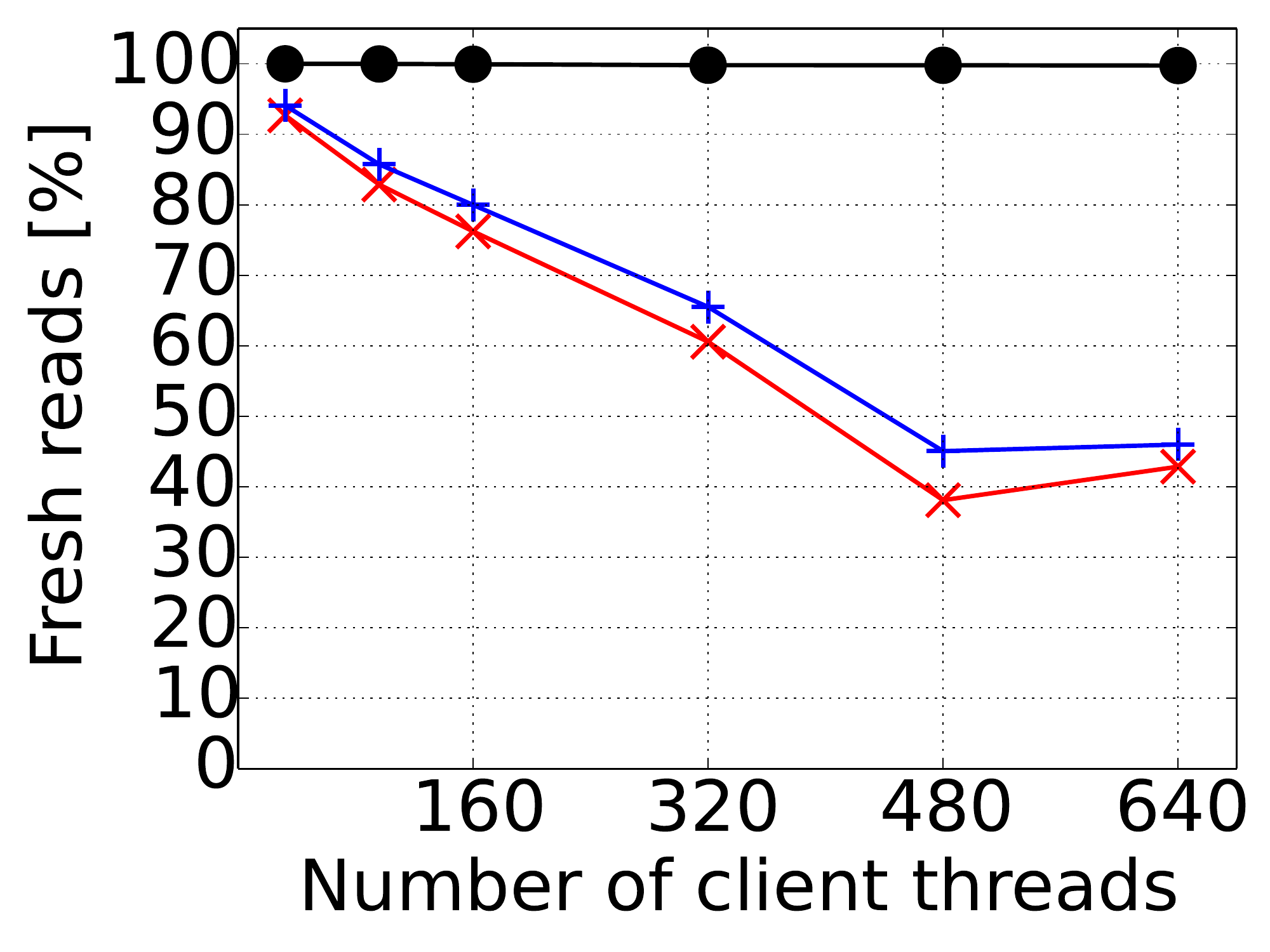}
  \caption{freshness w:1000}
  \label{fig:face-fresh-1000}
\end{subfigure}

\begin{subfigure}{0.32\columnwidth}
  \centering
  \includegraphics[width=\linewidth]{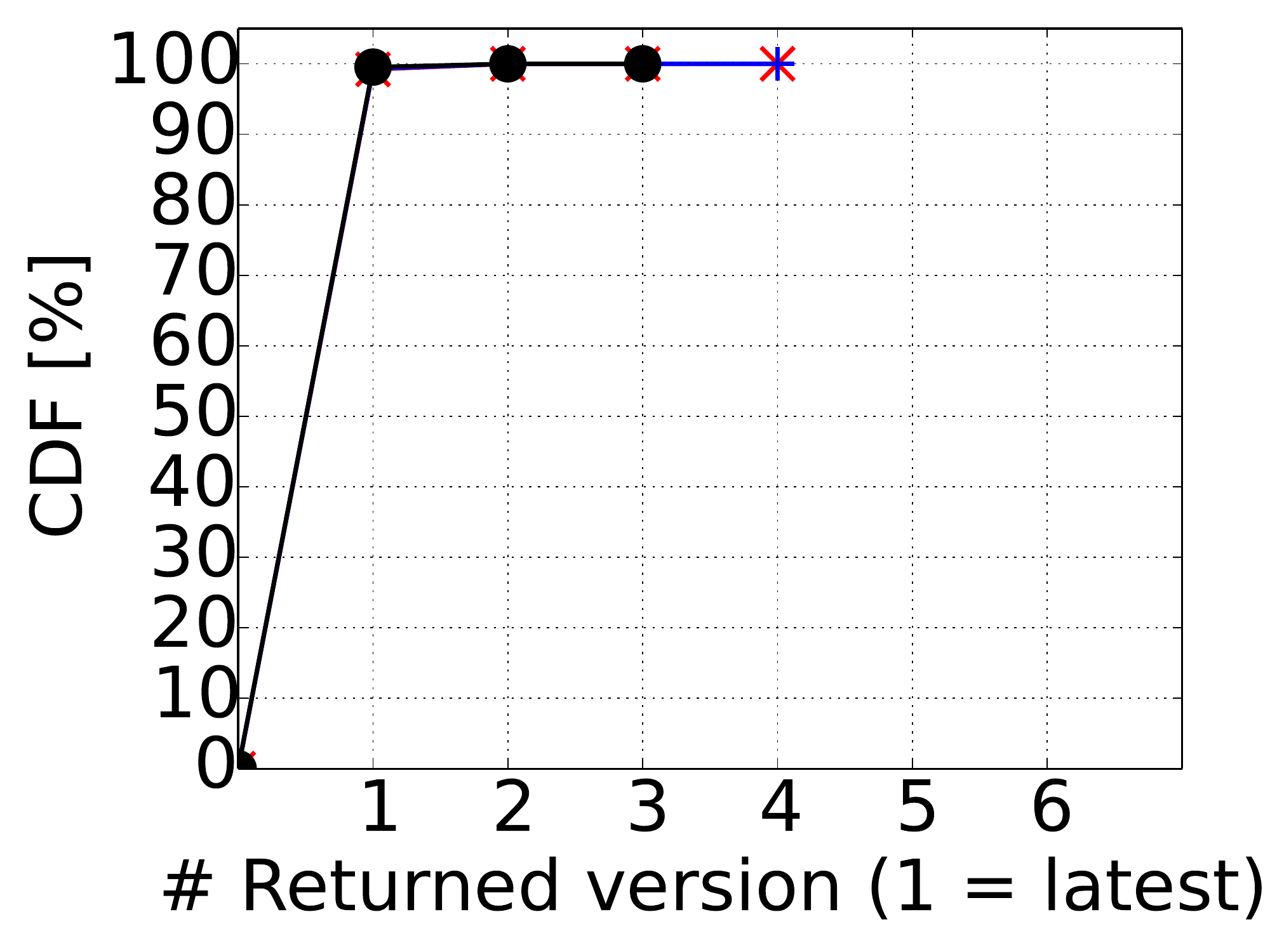}
  \caption{CDF w:2 th320}
  \label{fig:face-cdf-2}
\end{subfigure}
\begin{subfigure}{0.32\columnwidth}
  \centering
  \includegraphics[width=\linewidth]{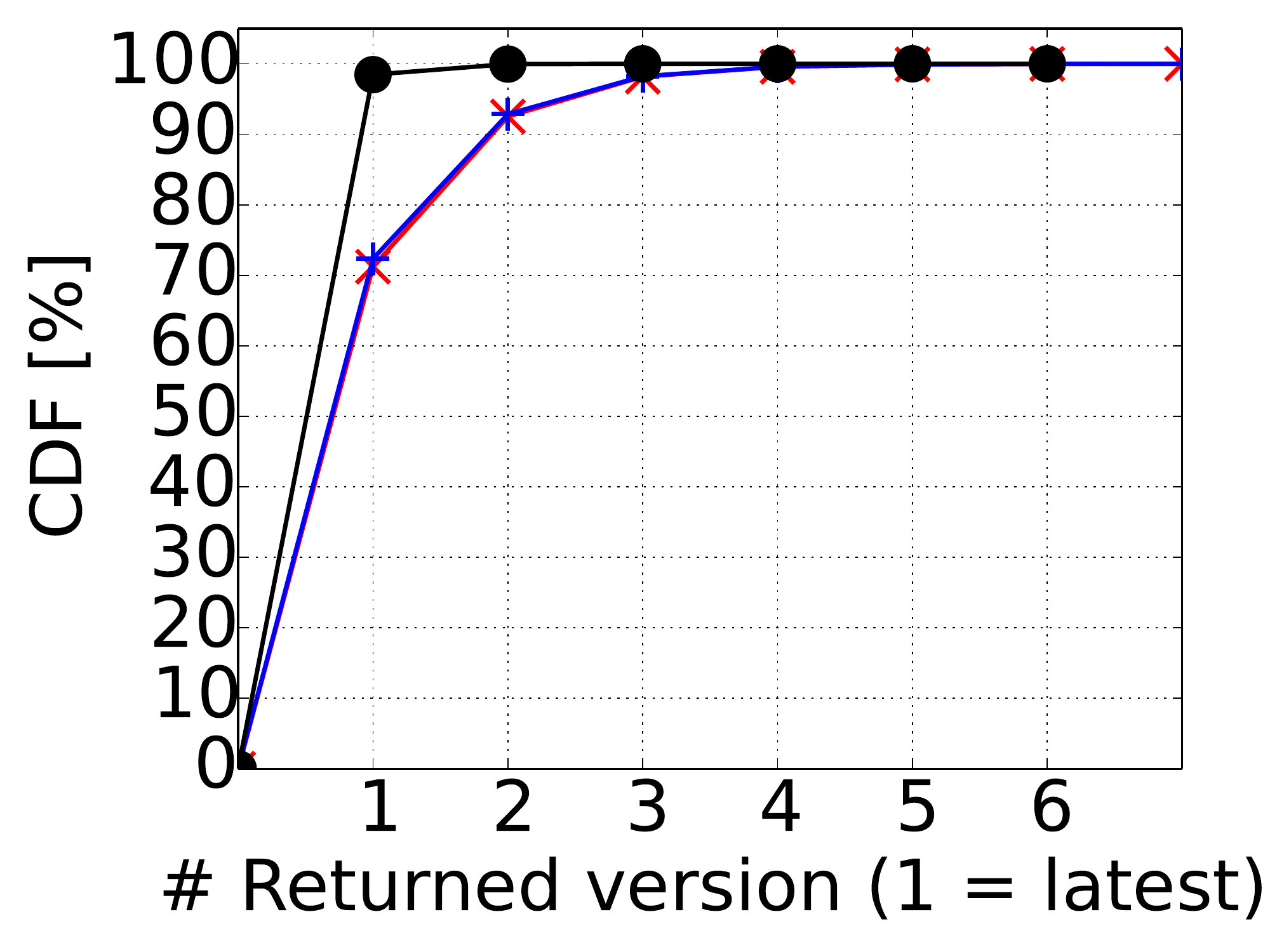}
  \caption{CDF w:100 th320}
  \label{fig:face-cdf-100}
\end{subfigure}
\begin{subfigure}{0.32\columnwidth}
  \centering
  \includegraphics[width=\linewidth]{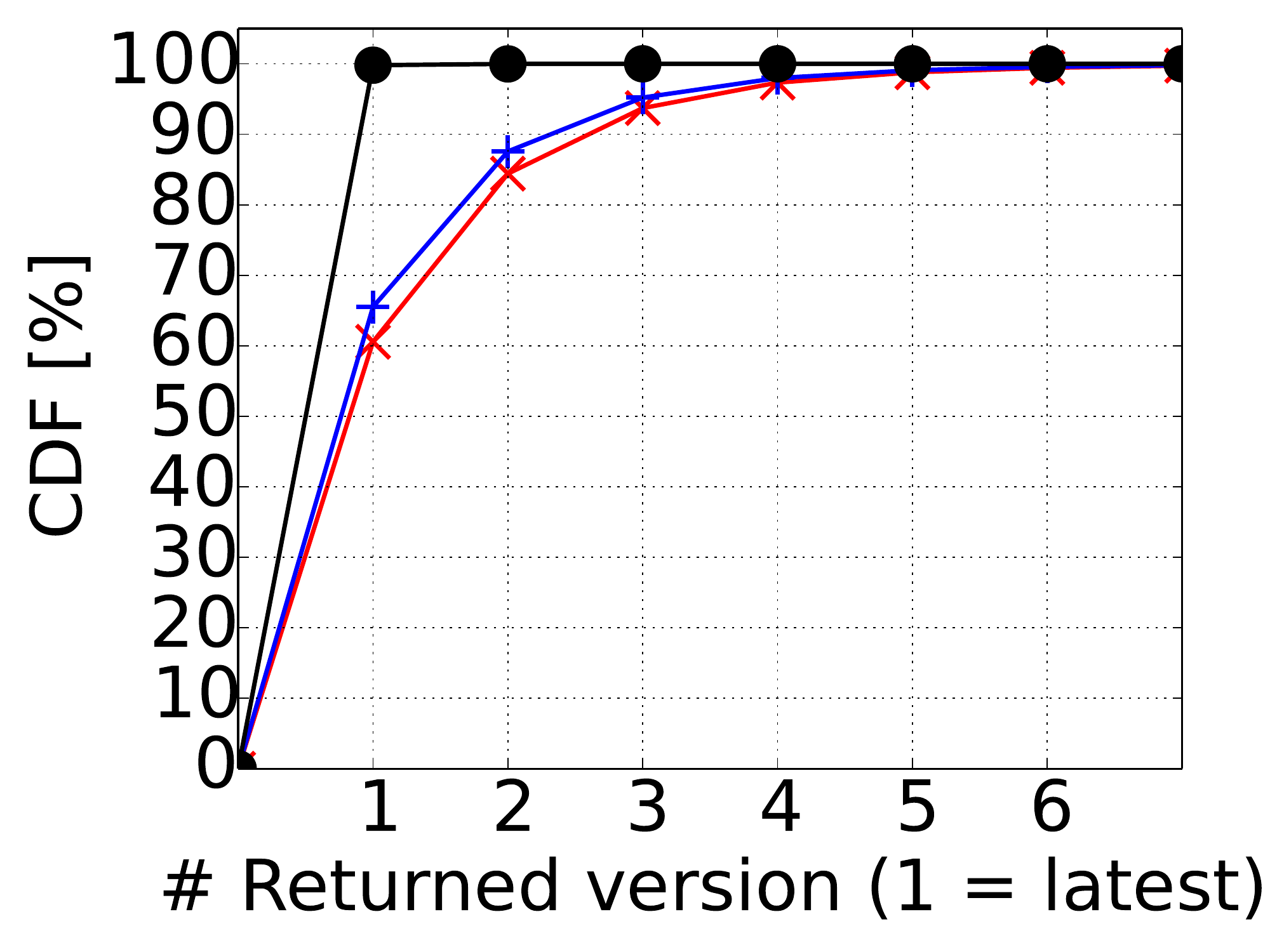}
  \caption{CDF w:1000 th320}
  \label{fig:face-cdf-1000}
\end{subfigure}

\begin{subfigure}{0.32\columnwidth}
  \centering
  \includegraphics[width=\linewidth]{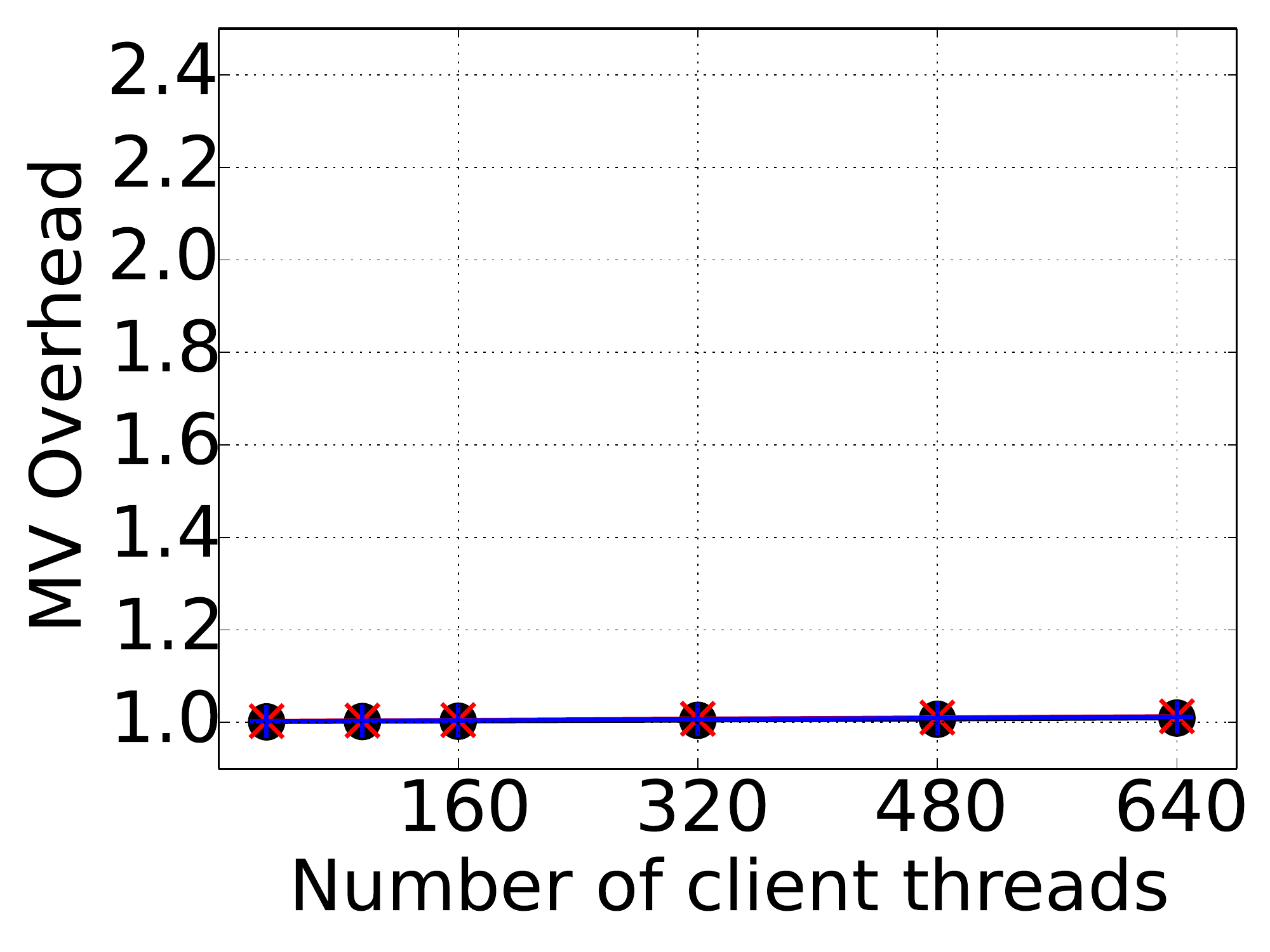}
  \caption{MV oh. w:2}
  \label{fig:face-moh-2}
\end{subfigure}
\begin{subfigure}{0.31\columnwidth}
  \centering
  \includegraphics[width=\linewidth]{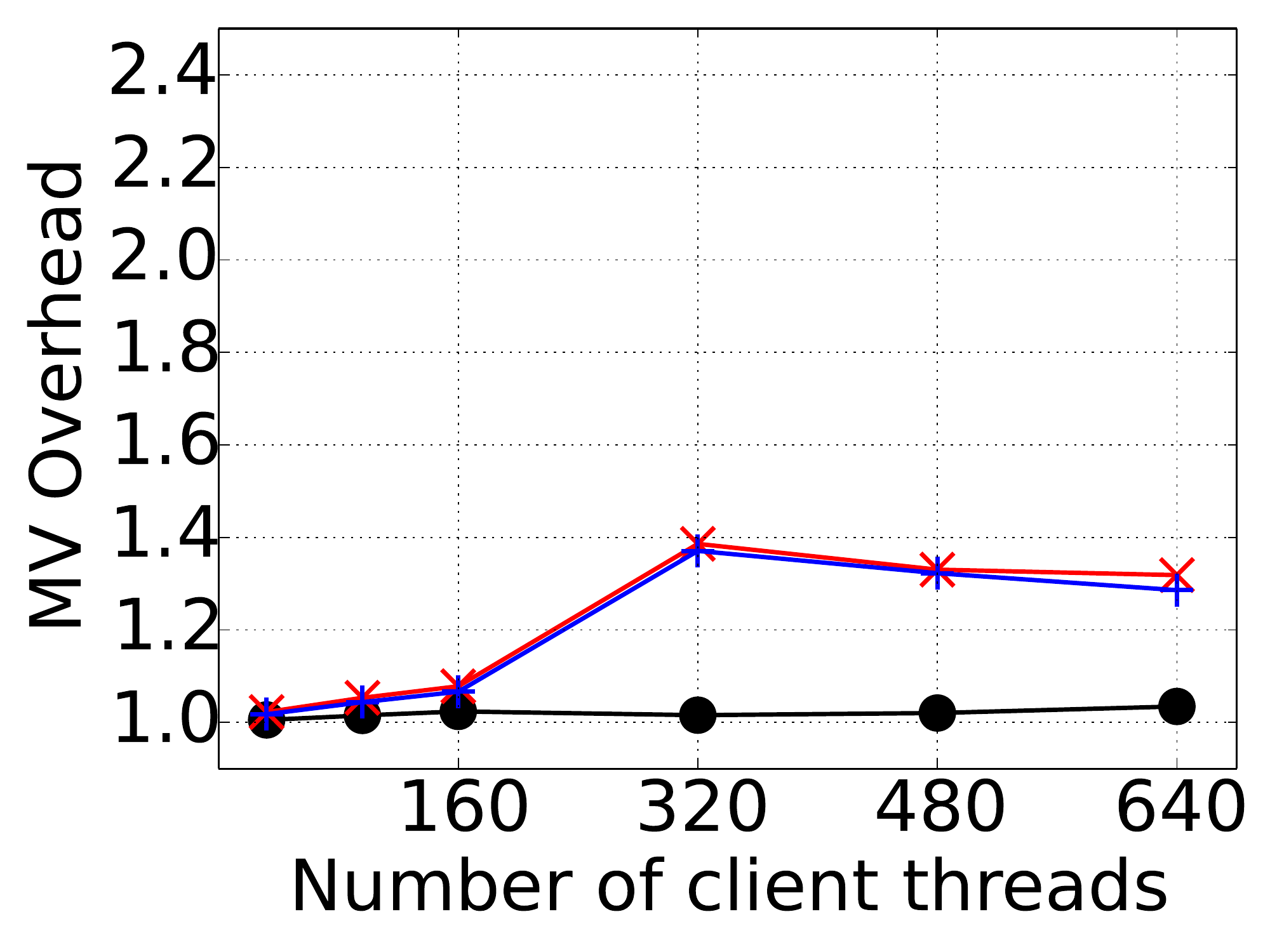}
  \caption{MV oh. w:100}
  \label{fig:face-moh-100}
\end{subfigure}
\begin{subfigure}{0.31\columnwidth}
  \centering
  \includegraphics[width=\linewidth]{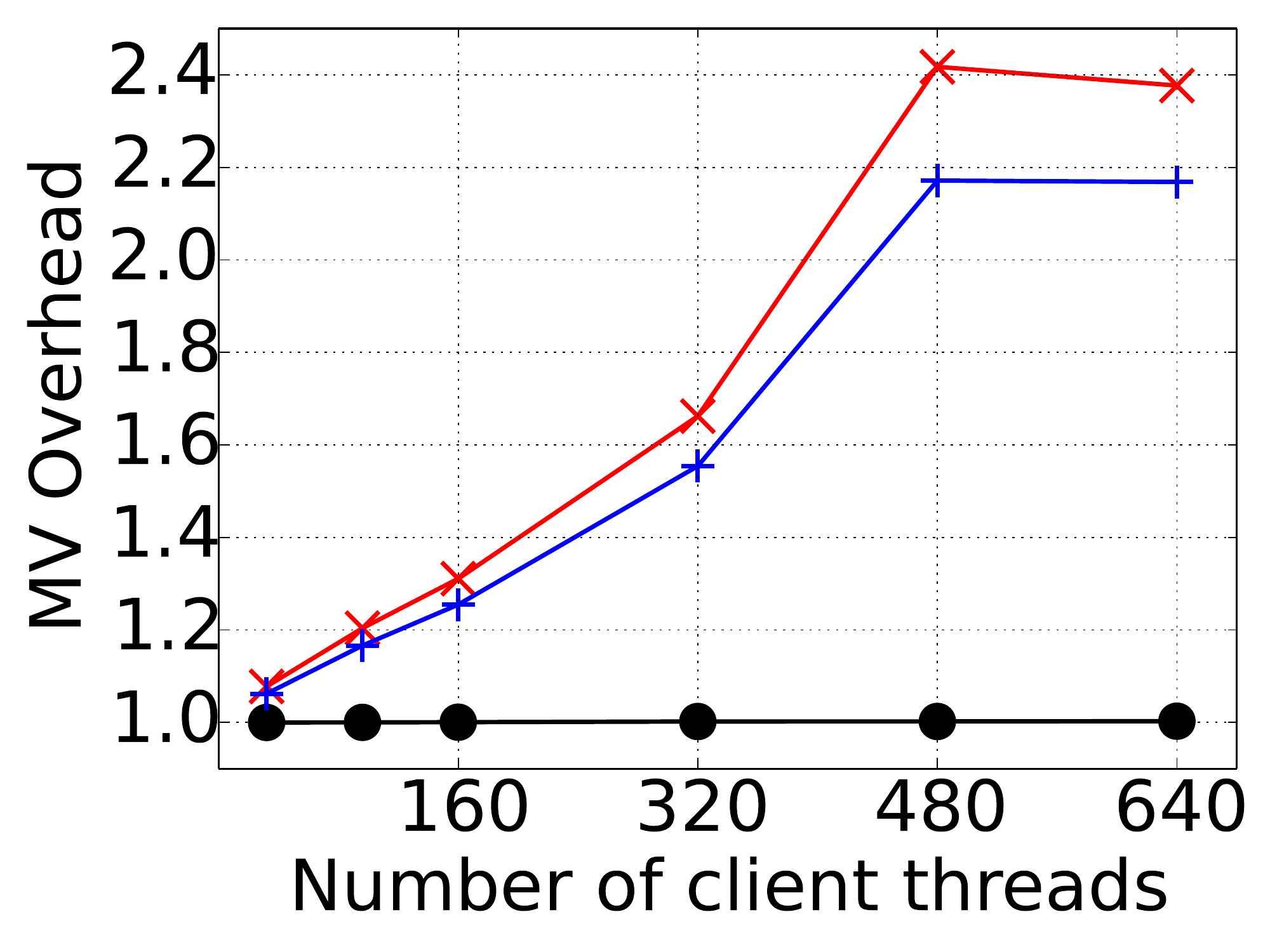}
  \caption{MV oh. w:1000}
  \label{fig:face-moh-1000}
\end{subfigure}
\caption{Multi-shot read-only transactions}
\label{fig:eval-face}
\end{figure}

We perform the same analysis under the multi-shot workload.
Figure \ref{fig:eval-face}
shows the results.
Results follow the same trend as those of the previous workload. 
However, some effects get diminished while others get augmented.
In what follows, we refer to the differences between results.

\smallskip\noindent\textbf{Latency.}
Figures \textbf{\ref{fig:one-lat-2}}, \textbf{\ref{fig:one-lat-10}} and \textbf{\ref{fig:one-lat-100}}
show the latency response of all protocols 
under this workload.
These transactions exhibit significantly higher ---around 10X--- latency than 
single-shot transactions, as they incur 10 rounds of 100 reads each.
The trend of all systems is very similar to that
of single-shot transactions: CV outperforms the remaining protocols, 
and the difference becomes larger as update rate augments.
One difference with respect to single-shot transactions is that 
OP exhibits slightly worst performance than the other systems.
This happens because of OP's mechanism for enforcing causal order:
every time a transaction coordinator receives a read response, 
it must recompute the transaction's causal dependencies
(Algorithm \ref{alg-coord}, Lines \ref{tcupdatemaxver} and \ref{tcupdatedepend}).
Under this workload, each transaction coordinator
performs this computation 1000 times.
Nevertheless, the protocol could be modified to avoid this situation
by performing such computation in parallel with subsequent read
operations, which we have not experimented with.

%
%
%

\smallskip\noindent\textbf{Freshness and multi-version overhead.}
Figures \textbf{\ref{fig:face-fresh-2}}, 
\textbf{\ref{fig:face-fresh-100}} and \textbf{\ref{fig:face-fresh-1000}} show the freshness response
as the number of client threads increases, for different update rates.
Figures \textbf{\ref{fig:face-cdf-2}}, \textbf{\ref{fig:face-cdf-100}} and \textbf{\ref{fig:face-cdf-1000}} 
display a CDF showing
how stale a read version is under 320 client threads for each update rate.
The trend is similar to that of single-shot read-only transactions: OP outperforms the remaining protocols
under all configurations, and Cure and AV degrade freshness significantly
as contention is added to the system.
For all protocols, the effect of staleness gets magnified with respect
to that of single-shot transactions.
This occurs because transactions are long lived, which renders interleaving with
update transactions more frequent.
The worst-case scenarios are 5\% of stale updates for OP, while 62\% for AV, and
55\% for Cure. 
In terms of oldest versions read under contention (50\% of updates and maximum client load),
OP returned, at most, the 7th-to-latest version,
while Cure the 28th and AV the 31st.
Figures \ref{fig:face-moh-2}, \ref{fig:face-moh-100},
and \ref{fig:face-moh-1000} show the multi-version-overhead
results under this workload.
Graphs follow a similar-but-magnified trend too as that of
single-shot reads, where overhead peaks at 1.05X for OP, 2.2X for Cure, and
2.4X for AV.

\smallskip\noindent\textbf{Conclusion.}
Under this workload, as transactions live long time, all protocols exhibit similar latency, including
Cure, which is not latency optimal.
In terms of freshness, we observe that protocols with Atomic Visibility get highly penalised
as contention increases.

\section{Related Work}
\label{sec:systems}
\smallskip\noindent\textbf{\emph{Impossibility results.}}
The CAP theorem proves that, in a replicated system, it is impossible
to guarantee strong consistency (C), high availability (A), and partition
tolerance (P) at the same time \cite{rep:pan:1628}.
Under partition, system designers must
decide between remaining available
but weakly-consistent (AP),
or remaining strongly-consistent but not available (CP).
Strong consistency requires, at the minimum,
ensuring that single object updates respect a total
monotonic order, which
requires synchronous replication.
This result has motivated designs that foster latency
to eschew both strong consistency and read isolation
\cite{facebook-challenges-consistency}, properties frequently mixed.
This effect has been exacerbated by the fact that
all existing AP designs that enforce isolation
exhibit delays.
This work shows that the performance of read isolation is
orthogonal to update-order enforcement.
Moreover, the presented algorithms are
available under partition and exhibit minimal delays.

Lu et al. prove that no protocol can achieve
Strict Serialisability (which implies Atomic Visibility)
and Minimal Delay \cite{snow}.
Therefore, their work proves that the upper-left 
corner of Figure \ref{fig:3-way-tradeoff} is unachievable under this model.  
Strict Serialisability requires that, 
if an update transaction $T_U$ commits updates at time $t_U$,
all transactions that start after $t_U$, in real time, observe 
$T_U$'s updates.
This requirement disallows that a transaction reads from an outdated snapshot.
Intuitively, knowing the latest set of committed transactions at a 
given point in time requires coordination, which implies added delay.
Their result is complementary to ours.

Didona et al. prove that a minimal-delay causally-consistent
(order-preserving) distributed read protocol forces expensive
updates \cite{didona18}. Their work considers a more-restrictive model;
a client, which coordinates a transaction, is not allowed to contact
a partition server (and vice-versa) outside the scope of the transaction.
This forbids a client to receive asynchronous notifications 
regarding stable snapshots.
This is instrumental to achieve Minimal Delay and progress in our protocols.
Without this restriction, that result is similar to
Proposition \ref{prop-impossible}.

Attiya et al. determined bounds on the response
times of Sequential Consistency and Linearisability, consistency guarantees of 
single objects \cite{attiya-welch}.
The results of that work can be used to help determining the delays of implementations that
rely on blocking to ensure a given isolation property 
(e.g., Strict Serialisability requires linearisable objects).

\smallskip\noindent\textbf{Systems.}
We relate the properties 
of existing systems to the trade-off.
Table~\ref{tab-relwork} summarises the discussion.

\begin{table}[!thp]
	\centering
  \includegraphics[width=1\linewidth]{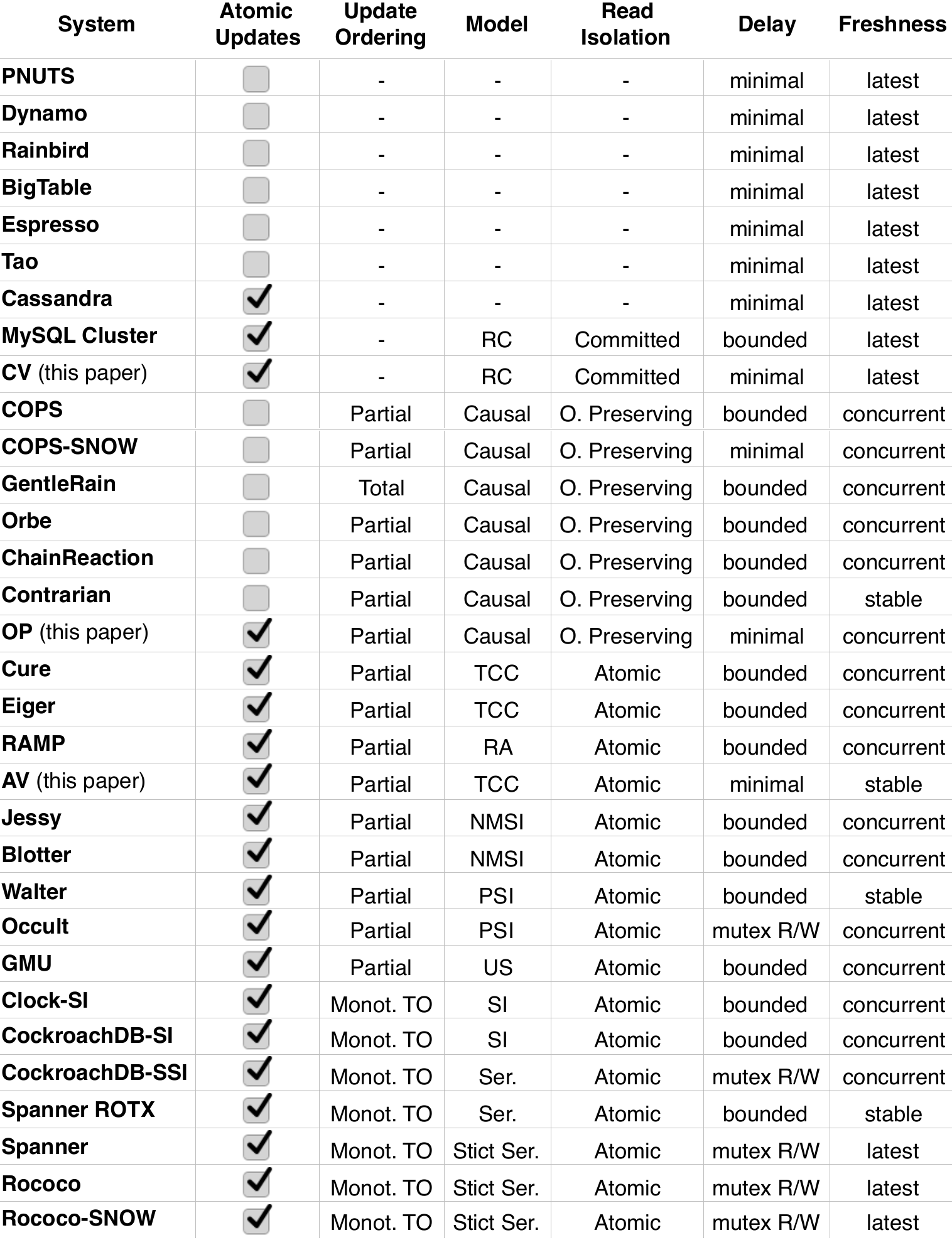}
    \caption{
      Guarantees, delay and freshness for several published systems.
      \rm
      %
  }
  \label{tab-relwork}
\end{table}

\smallskip\noindent\emph{Weak Isolation.}
Systems that are designed for high-availability and low latency under 
replication are those that do not require a per-object monotonic total order of
updates, and thus avoid synchronous replication. 
Surprisingly, minimal-delay designs with
read isolation are missing from the literature, 
as they all incur delays.

\noindent{No Isolation.}
Espresso \cite{linkedin-espresso},
Tao \cite{tao}, Yahoo's PNUTS \cite{rep:1685}, 
Amazon's Dynamo \cite{app:rep:optim:1606},
Twitter's Rainbird \cite{rainbird}, 
and Google's BigTable \cite{bigtable} ensure optimal reads but no atomic updates
or read isolation.
Cassandra offers atomic updates and no isolation.
Reads are only prevented from observing partial updates
within a row \cite{cassandra-isolation}.

\noindent{Committed Visibility.}
In MySQL cluster, reads may block waiting for a transaction to commit \cite{mysql-cluster}. 
We introduce CV, a protocol providing atomic updates
and optimal committed reads.

\noindent{Order-Preserving Visibility.}
All systems we discuss provide snapshot reads
that preserve causal order. 
None supports all-or-nothing updates. 
Similarly to Cure \cite{rep:pro:sh182},
GentleRain \cite{db:syn:1752}, Orbe \cite{rep:alg:1725}
and ChainReaction \cite{syn:pan:1750}
block to wait for concurrent transactions to commit
and for clocks to catch up.
COPS \cite{rep:syn:1662} incurs multiple read-rounds.
In Contrarian \cite{didona18}, a transaction executes
a communication round to any partition server to obtain
a stable snapshot, followed by the read phase.
COPS-SNOW \cite{snow} offers Minimal Delay and
Concurrent Freshness.
It removes the second round of reads in COPS by rendering updates expensive.
An update operation must update data structures of all 
the objects it causally depends on.
Both of these systems rely on metadata sized with the 
number of objects in the system to causally-order updates.
The introduced OP has the same read semantics without incurring
such costs, and furthermore providing atomic updates.

\noindent{Atomic Visibility.}
Cure exhibits the blocking scenarios introduced in Section \ref{base-protocol}.
The remaining considered systems incur multiple rounds of reads.
Examples include Eiger \cite{syn:rep:1708}
and RAMP \cite{ramp}.
This work introduced AV, the first weakly-consistent 
protocol that achieves minimal delays and Atomic 
Visibility. It implements TCC.

\smallskip\noindent\emph{Strong Isolation.}
Cassandra offers single-object strong consistency through synchronous replication,
and no cross-object isolation \cite{cassandra-consistency}.

Walter, Occult, Blotter,
Jessy and GMU ensure causal order 
and enforce a per-object total order
by avoiding conflicting writes.
They achieve Concurrent Freshness with delays. 
Walter retries reads. Blotter, Jessy and GMU read sequentially.
Occult exhibits Unbounded Delays. 
A transaction attempts to read from an
atomic snapshot from sites that might be in an
inconsistent state. It aborts when it detects an inconsistency.

Clock-SI provides Snapshot Isolation.
Its read algorithm is very similar to that of Cure and GentleRain;
it blocks in the case of clock skew or waiting for transactions to commit.
CockRoachDB offers Snapshot Isolation and
Serialisability, both ensuring Atomic Visibility.
Under the former, reads might block waiting for a transaction
to commit.
Under the latter, a read-only transaction might abort (indefinitely) when 
it detects a serialisation conflict.

Spanner features two kinds of transactions.
Strictly-serialisable transactions rely on locks to ensure
mutually-exclusive reads and writes (by Proposition 
\ref{prop-latest-freshness}, this is unavoidable).
Spanner's read-only transactions exhibit
Bounded Delay as a server might need to wait for
physical clocks to advance past a transaction's snapshot time.
Rococo provides Strict Serialisability (Atomic Visibility) under
Latest Freshness.
Its read algorithm issues an unbounded number of rounds to 
ensure its desirable guarantees.
Rococo-SNOW issues a bounded number of read rounds, and
blocks updates when these rounds do not attain Atomic Visibility
and Latest Freshness. Once updates are stopped, these are guaranteed.

\section{Conclusion}

We have explored the three-way trade-off between 
a transactional read algorithm's isolation guarantees, its delays, and its freshness,
and analyse a spectrum of possible points in the design space.
Interestingly, order-preserving minimal-delay reads can be fresher than (the strongest) atomic.
Moreover, reading the most up-to-date data and ensuring 
isolated reads, required by strict serialisability,
is only possible by implementing mutually-exclusive reads
and updates, which may delay reads indefinitely.
We have used these results to guide protocol design.
Departing from an existing transactional protocol exhibiting delays, 
we have created three minimal-delay variants: 
one maintains its read guarantees by degrading its freshness, 
while the other two improve freshness in different degrees 
by degrading read guarantees.
The evaluation of these three protocols supports 
the theoretical conclusions of the trade-off.

\section*{Acknowledgments}
This research is supported in part 
  \begin{inparablank}
    \item
    by European FP7 project
    \href{http://syncfree.lip6.fr/}{609\,551 SyncFree} (2013--2016),
    \item
    by European H2020 project
    \href{http://LightKone.eu/}{LightKone \#732\,505} (2017--2020), and
    \item
      by the \href{http://RainbowFS.lip6.fr/}{RainbowFS} project
      of \emph{Agence Nationale de la Recherche}, France, number 
      ANR-16-CE25-0013-01.
  \end{inparablank}

\bibliographystyle{plainnat}
\bibliography{predef,references1,shapiro-bib-ext,shapiro-bib-perso}

\appendix
\section{Protocols: session guarantees}
\label{app:sesgar}

\noindent\textbf{Read your writes.}
The algorithms presented in Section \ref{sec-protocols}
do not ensure 
the \emph{read your writes} session guarantee, required by causal consistency.
The problem arises because updates 
must undergo the stabilisation process to be available to further transactions.
If a client's transaction performs some updates, 
a subsequent transaction by the same client might miss
its latest updates because they are not stable yet.

Read your writes can be enforced by a client caching its
latest updates.
When receiving a read response, the client
compares the version received with the cached one (if any).
If the cached version is fresher, the one returned by the system
is discarded.
After a transaction finishes, a TC returns
the latest stable vector $SV$ it is currently aware of.
A client can invalidate all updates with $cv \leq SV$.  

\smallskip\noindent\textbf{Monotonic Reads.}
Under AV and OP, monotonic reads are ensured if a client always connects
to the same server.
However, if the server fails or becomes unreachable, 
a client might connect to a server where $SV$ is behind that of 
the $ss$ of the client's last transaction. 
This might lead to observing less up-to-date data
than what the client previously observed.
This violates the monotonic-reads session guarantee.

To ensure monotonic reads, a client informs
a TC of its latest $ss$.
When a transaction commits, the TC returns
the transaction's $ss$.
If a TC detects that its $SV$ is behind the client's $ss$ ($SV < ss$), 
it catches up as follows:
if the TC runs at the same site as the client's latest transaction,
it updates its $SV$ to $ss$ immediately.
This is possible because the snapshot
was previously observed as stable by another partition at the same
site.
If the TC runs at a different site, TC has to block
until the stabilisation protocol validates $SV \geq ss$.

\section{Evaluation of blocking in Cure}
\label{app:cure-blocking}
\begin{figure}[t!]
\centering
	\includegraphics[width=.6\linewidth]{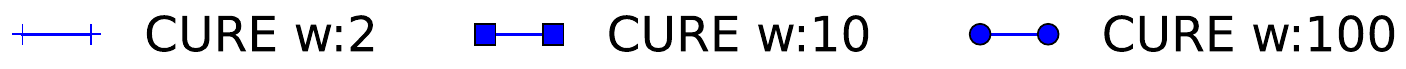}
    \begin{subfigure}{0.5\columnwidth}
  \centering
  \includegraphics[width=.8\linewidth]{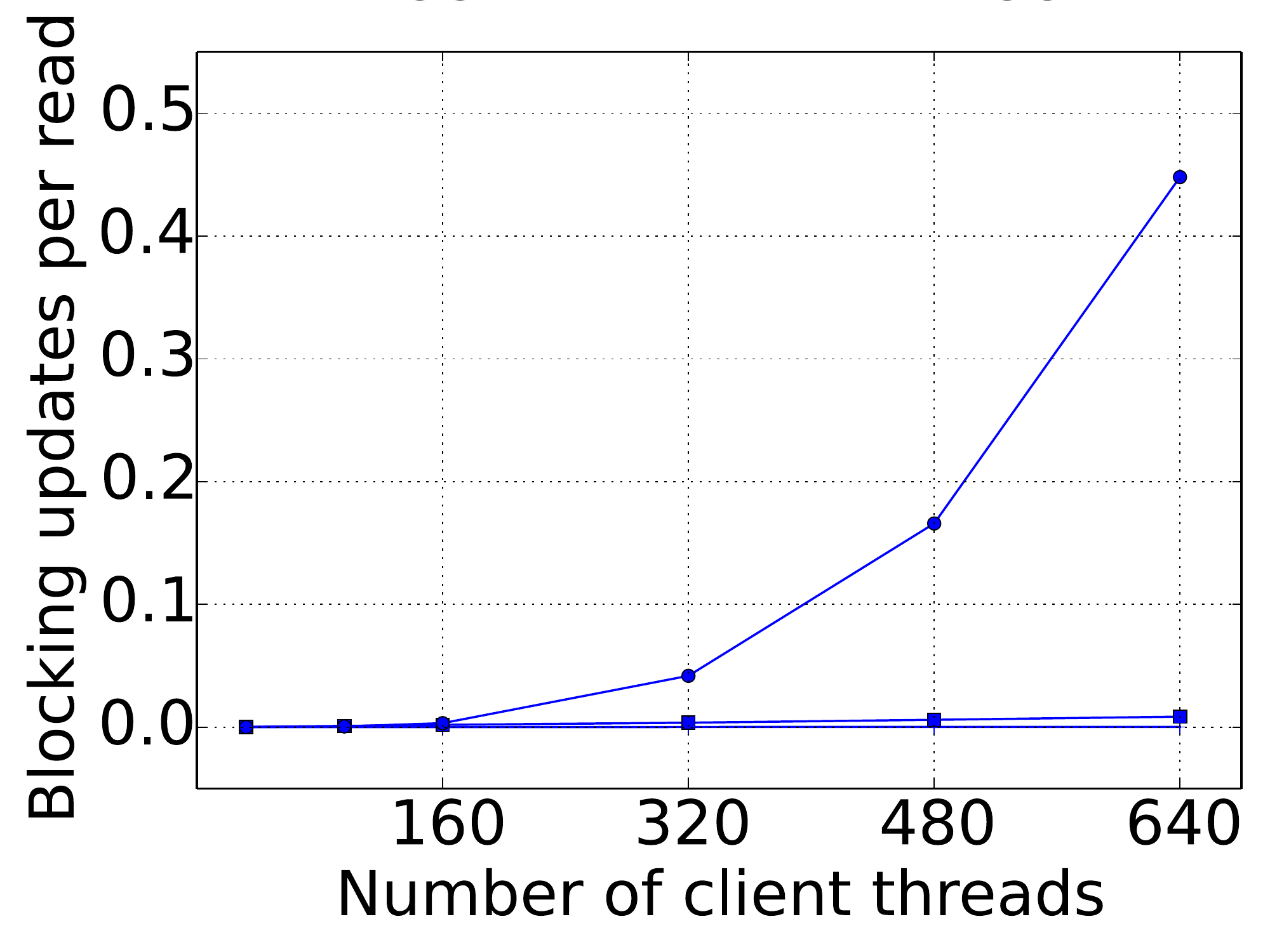}
  \caption{Wait due to committing update}
  \label{fig:single-cure-skew}
\end{subfigure}%
\begin{subfigure}{0.5\columnwidth}
\centering
  \includegraphics[width=.8\linewidth]{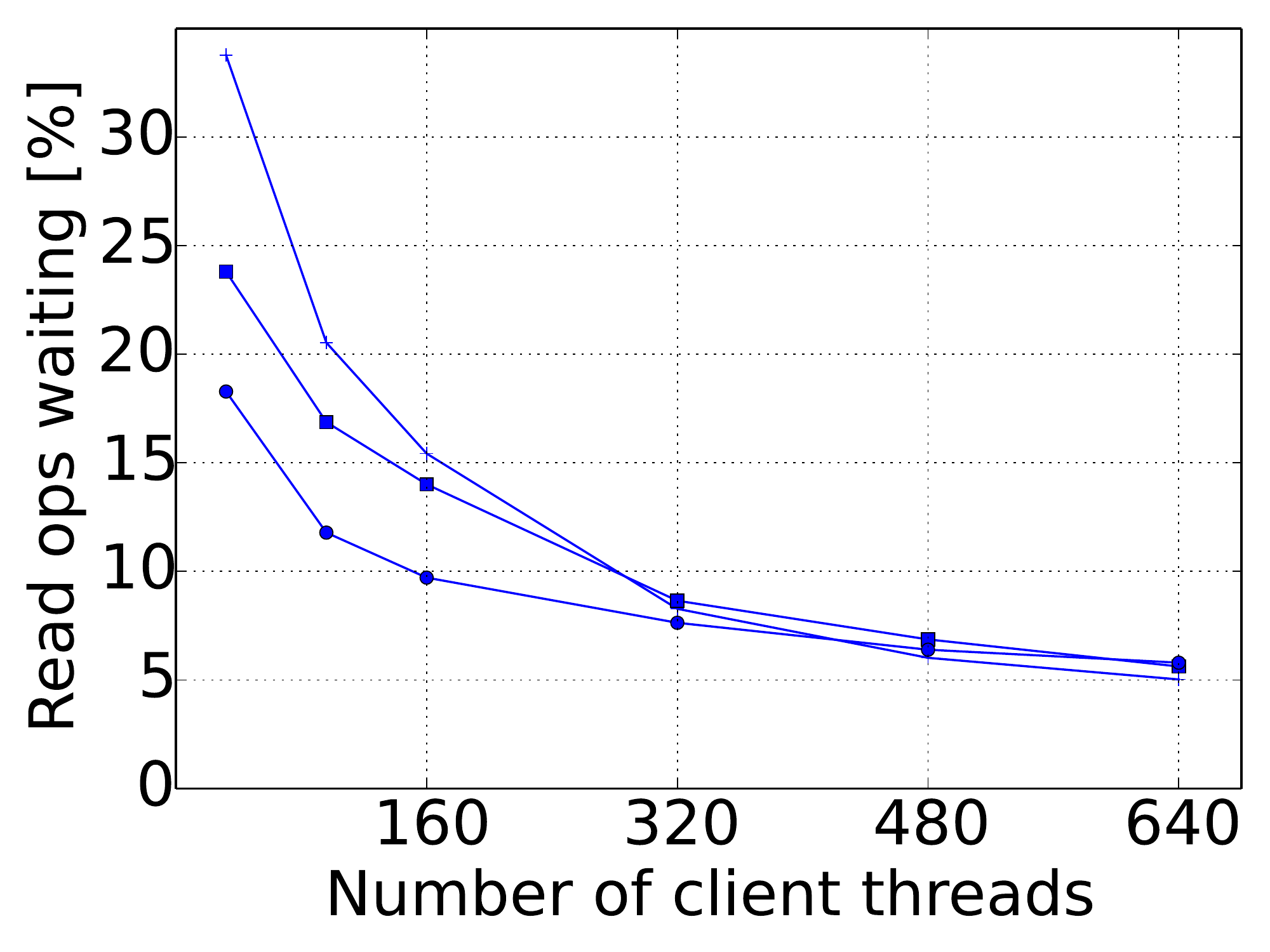}
  \caption{Wait due to clock skew}
  \label{fig:single-cure-update}
\end{subfigure}

\caption{Cure blocking scenarios, single-shot workload}
\label{fig:cure-single}
\end{figure}

\noindent\textbf{Single-shot RO transactions.}
The design of Cure is not latency optimal.
We plot, in Figure \textbf{\ref{fig:single-cure-update}}, the percentage of read operations that blocked
due to clock skew.
This effect is frequent under small number
of client threads, and dissipates as the system becomes more loaded.
Under high load, the time it takes to process a received read-request message is larger, and during
this time, lagging clocks can catch up.
Figure \textbf{\ref{fig:single-cure-skew}} shows how, when keys become highly contended (i.e., at high update rate
and number of client threads), waiting for an update operation becomes frequent.
At maximum contention ---640 threads and 50\% of updates--- a read operation waits, on average,
for 0.45 update operations to finish or, equivalently, each transaction waits for an average of
45 updates to commit.

\begin{figure}[t!]
\centering
	\includegraphics[width=.6\linewidth]{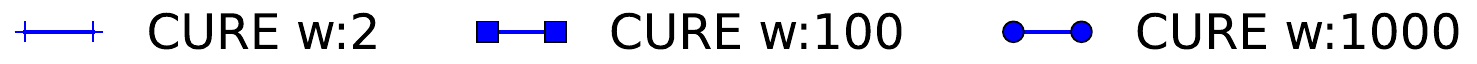}
    \begin{subfigure}{0.5\columnwidth}
  \centering
  \includegraphics[width=0.8\linewidth]{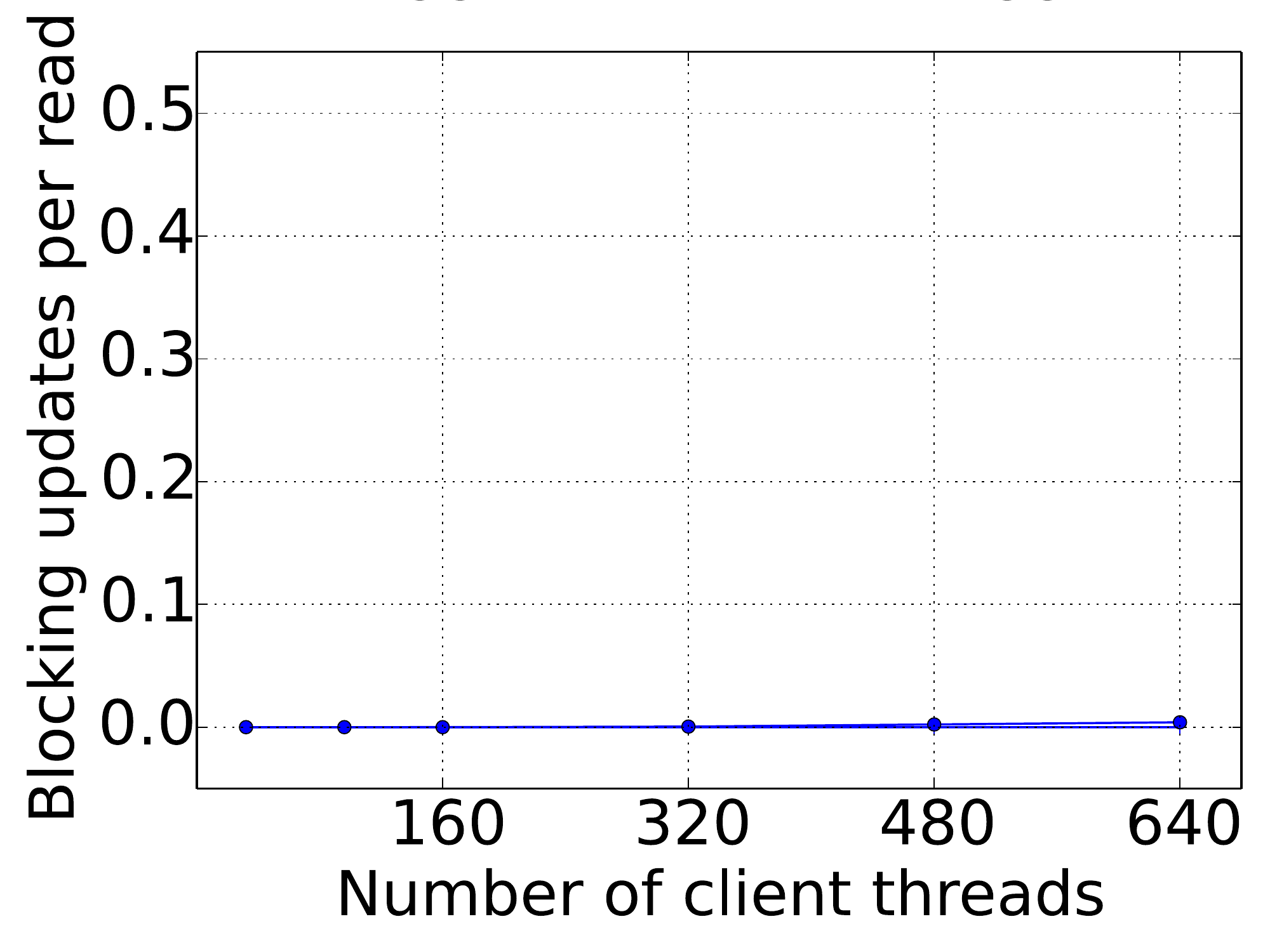}
  \caption{Wait due to committing update}
  \label{fig:face-cure-skew}
\end{subfigure}%
\begin{subfigure}{0.5\columnwidth}
\centering
  \includegraphics[width=0.8\linewidth]{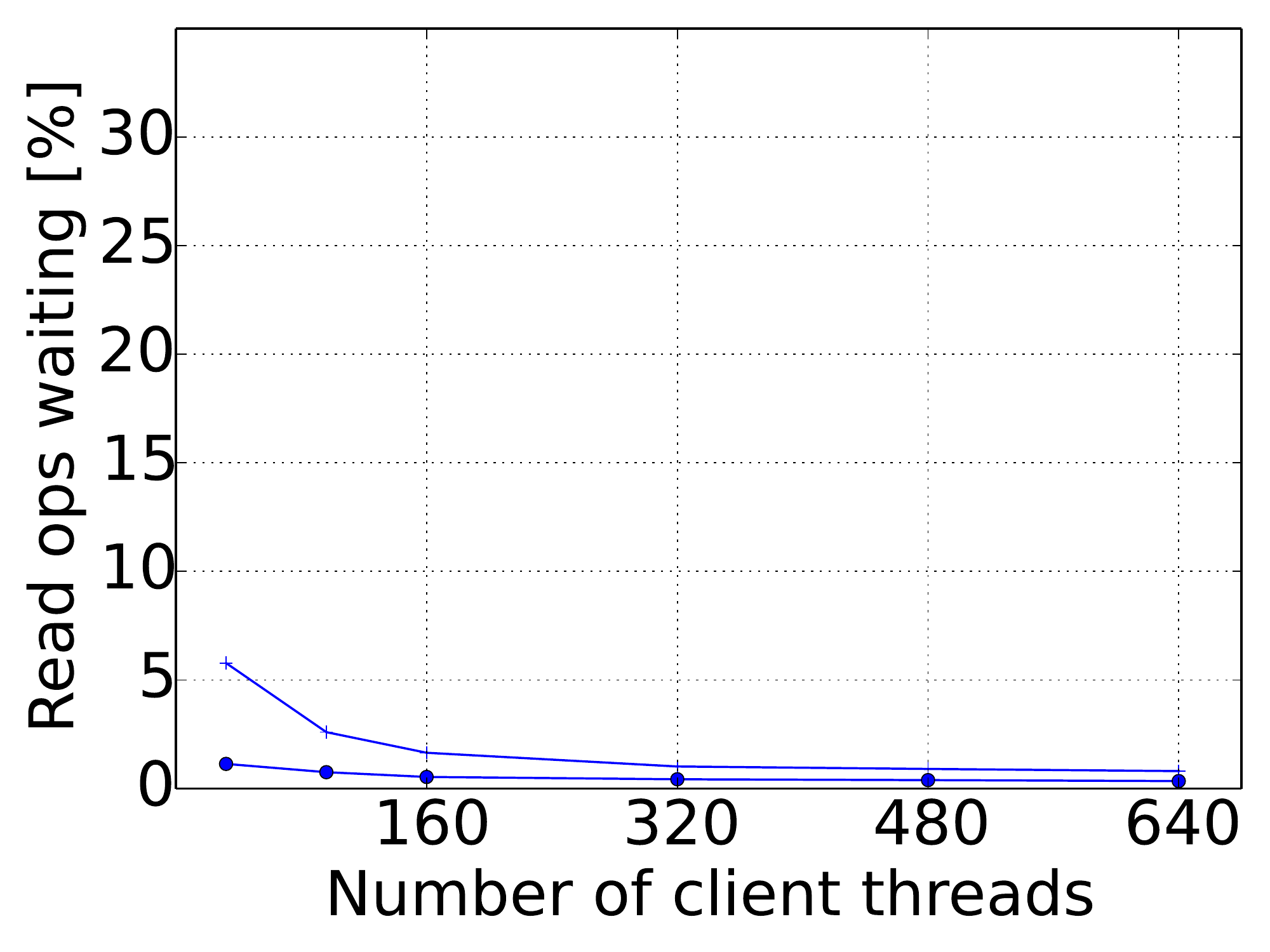}
  \caption{Wait due to clock skew}
  \label{fig:face-cure-update}
\end{subfigure}

\caption{Cure blocking scenarios, multi-shot workload}
\label{fig:cure-face}
\end{figure}

\smallskip\noindent\textbf{Multi-shot RO transactions.}
Under this workload, where transactions execute for a long time,
the blocking cases of Cure are significantly reduced with respect to
those of single-shot transactions.
Figure \textbf{\ref{fig:face-cure-skew}} shows the percentage of read operations that blocked
due to clock skew under Cure. As we see, the effect practically disappears
---below 6\% of reads block--- under all workloads.
If we consider that each read round takes approximately 10ms, rounds after the first one are very
unlikely to block due to clock skew, where most of waiting is expected to happen.
The same occurs with blocking due waiting for update transactions to commit, as shown in
Figure \textbf{\ref{fig:face-cure-update}}. 
Under maximum contention, under 1\% of read operations
block due to this effect.

\end{document}